\documentclass[twoside,twocolumn,9pt]{article}
\usepackage{extsizes}
\usepackage[super,sort&compress,comma]{natbib} 
\usepackage[version=3]{mhchem}
\usepackage[left=1.5cm, right=1.5cm, top=1.785cm, bottom=2.0cm]{geometry}
\usepackage{balance}
\usepackage{mathptmx}
\usepackage{sectsty}
\usepackage{graphicx} 
\usepackage{lastpage}
\usepackage[format=plain,justification=justified,singlelinecheck=false,font={stretch=1.125,small,sf},labelfont=bf,labelsep=space]{caption}
\usepackage{float}
\usepackage{fancyhdr}
\usepackage{fnpos}
\usepackage[english]{babel}
\addto{\captionsenglish}{%
  
}
\usepackage{array}
\usepackage{droidsans}
\usepackage{charter}
\usepackage[T1]{fontenc}
\usepackage[usenames,dvipsnames]{xcolor}
\usepackage{setspace}
\usepackage[compact]{titlesec}
\usepackage{hyperref}
\usepackage{subcaption}
\usepackage{mathtools, cuted}
\usepackage{pdfpages} 
\usepackage{lipsum,widetext}

\definecolor{cream}{RGB}{222,217,201}

\begin{document}

\pagestyle{fancy}
\thispagestyle{plain}
\fancypagestyle{plain}{
\renewcommand{\headrulewidth}{0pt}
}

\makeFNbottom
\makeatletter
\renewcommand\LARGE{\@setfontsize\LARGE{15pt}{17}}
\renewcommand\Large{\@setfontsize\Large{12pt}{14}}
\renewcommand\large{\@setfontsize\large{10pt}{12}}
\renewcommand\footnotesize{\@setfontsize\footnotesize{7pt}{10}}
\makeatother

\renewcommand{\thefootnote}{\fnsymbol{footnote}}
\renewcommand\footnoterule{\vspace*{1pt}%
\color{cream}\hrule width 3.5in height 0.4pt \color{black}\vspace*{5pt}} 
\setcounter{secnumdepth}{5}

\makeatletter 
\renewcommand\@biblabel[1]{#1}            
\renewcommand\@makefntext[1]%
{\noindent\makebox[0pt][r]{\@thefnmark\,}#1}
\makeatother 
\renewcommand{\figurename}{\small{Fig.}~}
\sectionfont{\sffamily\Large}
\subsectionfont{\normalsize}
\subsubsectionfont{\bf}
\setstretch{1.125} 
\setlength{\skip\footins}{0.8cm}
\setlength{\footnotesep}{0.25cm}
\setlength{\jot}{10pt}
\titlespacing*{\section}{0pt}{4pt}{4pt}
\titlespacing*{\subsection}{0pt}{15pt}{1pt}

\fancyfoot{}
\fancyfoot[RO]{\footnotesize{\sffamily{\thepage}}}
\fancyfoot[LE]{\footnotesize{\sffamily{\thepage}}}
\fancyhead{}
\renewcommand{\headrulewidth}{0pt} 
\renewcommand{\footrulewidth}{0pt}
\setlength{\arrayrulewidth}{1pt}
\setlength{\columnsep}{6.5mm}
\setlength\bibsep{1pt}

\makeatletter 
\newlength{\figrulesep} 
\setlength{\figrulesep}{0.5\textfloatsep} 

\newcommand{\topfigrule}{\vspace*{-1pt}%
\noindent{\color{cream}\rule[-\figrulesep]{\columnwidth}{1.5pt}} }

\newcommand{\botfigrule}{\vspace*{-2pt}%
\noindent{\color{cream}\rule[\figrulesep]{\columnwidth}{1.5pt}} }

\newcommand{\dblfigrule}{\vspace*{-1pt}%
\noindent{\color{cream}\rule[-\figrulesep]{\textwidth}{1.5pt}} }

\makeatother

\twocolumn[
  \begin{@twocolumnfalse}
\vspace{1em}
\sffamily

\noindent\LARGE{\textbf{A phase field model combined with genetic algorithm for polycrystalline hafnium zirconium oxide ferroelectrics}} \\

\noindent\large{Sandeep Sugathan,\textit{$^{a}$} Krishnamohan Thekkepat,\textit{$^{b,c}$}, Soumya Bandyopadhyay, \textit{$^{a}$} Jiyoung Kim, \textit{$^{d}$} and Pil-Ryung Cha $^{\ast}$\textit{$^{a}$}} \\

\noindent\normalsize{Ferroelectric hafnium zirconium oxide (HZO) thin films show significant promise for applications in ferroelectric random-access memory, ferroelectric field-effect transistors, and ferroelectric tunneling junctions. However, there are shortcomings in understanding ferroelectric switching, which is crucial in the operation of these devices. Here a computational model based on phase field method is developed to simulate the switching behavior of polycrystalline HZO thin films. Furthermore, we introduce a novel approach to optimize the effective Landau coefficients describing the free energy of HZO by combining the phase field model with a genetic algorithm. We validate the model by accurately simulating switching curves for HZO thin films with different ferroelectric phase fractions. The simulated domain dynamics during switching also shows amazing similarity to the available experimental observations. The present work also provides fundamental insights into enhancing the ferroelectricity in HZO thin films by controlling grain morphology and crystalline texture. It can potentially be extended to improve the ferroelectric properties of other hafnia based thin films. 
} \\


 \end{@twocolumnfalse} \vspace{0.6cm}

  ]

\renewcommand*\rmdefault{bch}\normalfont\upshape
\rmfamily
\section*{}
\vspace{-1cm}


\footnotetext{\textit{$^{a}$~School of Advanced Materials Engineering, Kookmin University, Seoul 02707, Republic of Korea. E-mail: cprdream@kookmin.ac.kr}}
\footnotetext{\textit{$^{b}$~Electronic Materials Research Center, Korea Institute of Science and Technology,
Seoul 02792, Republic of Korea.}}
\footnotetext{\textit{$^{c}$~Division of Nano $\&$ Information Technology, KIST School, Korea University of Science and Technology, Seoul 02792, Republic of Korea.}}
\footnotetext{\textit{$^{d}$~Department of Materials Science and Engineering, The University of Texas at Dallas, 800 West Campbell Road, Richardson, Texas 75080, USA.}}


\section{Introduction}
\medskip

Reports of ferroelectricity in hafnium zirconium oxide (HZO) thin films~\cite{muller2011ferroelectric,HZO1} attracted considerable attention from the energy conversion and non-volatile memory devices community, which mainly focused on perovskite based materials. Hafnia based ferroelectric devices are preferable over perovskites for their complementary metal–oxide–semiconductor compatibility, back-end-of-line compatibility, and atomic layer deposition capability~\cite{park2015ferroelectricity,park2018review,mikolajick2018ferroelectric,kim2019ferroelectric,lee2020scale,cheema2020enhanced}. The origin of ferroelectricity has been studied extensively due to the formation of an otherwise unstable, polar orthorhombic phase, stabilized by dopants, defects, and kinetic factors during growth~\cite{huan2014pathways,sang2015structural,grimley2018atomic,lee2019nucleation,park2019thermodynamic,kunneth2019thermodynamics,xu2021kinetically,lee2021domains}. The polarization switching behavior of ferroelectric materials is a decisive factor in the operation of ferroelectric devices. The coercive field ($E_c$) and remnant polarization ($P_r$) of a ferroelectric material are the defining parameters for its hysteretic  behavior. In hafnia based thin films, the remnant polarization is comparable to that of conventional ferroelectrics, whereas the coercive field is much higher~\cite{boscke2011ferroelectricity,park2015ferroelectricity,hoffmann2016direct,kim2017large}. Indeed, the higher coercive field for hafnia based ferroelectrics is a limitation in ferroelectric random-access memory, but an advantage in ferroelectric field-effect transistor because of the superior polarization retention. It is essential to understand and model the switching dynamics in hafnia based ferroelectric devices since their performance can be significantly improved by optimizing coercive field and remnant polarization. 
\medskip

Phenomenological models based on Landau theory of phase transitions have been typically used to simulate domain dynamics and polarization switching in ferroelectrics. The phase field method based on continuum phenomenological description has been established as a powerful computational tool for understanding domain structures and switching behavior in ferroelectrics. In phase field model of ferroelectrics, the domain evolution is driven by the reduction in the total free energy of an inhomogeneous domain structure including the chemical driving force, domain wall energy, electrostatic energy as well as elastic energy~\cite{chen2008phase,wang2019understanding,wang2020strain}. The free energy is described in terms of Landau expansion coefficients which can be obtained from experimental data, first-principles calculations, or microscopic calculations~\cite{wang2013mechanics,pitike2019landau}. Unlike the extensively investigated conventional perovskites, for which the coefficients are
available in the literature, there is a lack of in-depth research on the derivation of Landau coefficients for hafnia based ferroelectrics. 
\medskip

In recent computational studies on ferroelectric HZO, the Landau coefficients are extracted by calibrating the intrinsic polarization hysteresis function, $P(E)$ with polarization-electric field (PE) curves obtained from experiments~\cite{hoffmann2016direct,khan2015negative,noh2019switching,hoffmann2019unveiling,saha2019phase,saha2020multi}. The switching characteristics simulated using the extracted Landau coefficients in many studies have shown dissimilarities with the measured data. Our goal is to efficiently estimate the Landau coefficients from the measured PE curve, such that the discrepancies in the simulated hysteresis curve are minimized. Therefore, in this study, we develop a simplified polycrystalline ferroelectric phase field model considering only $180^\circ$ ferroelectric domains and neglecting the elastic and depolarizing energy contributions to reduce the computational complexity. We combine this phase field model with a genetic algorithm (GA) to predict effective Landau coefficients for a {Hf$_{0.5}$Zr$_{0.5}$O$_2$} thin film. Genetic algorithms are randomized searching methods that converge to a global minimum of problem specific objective functions~\cite{Sastry2005}. The single-domain ground state property of the HZO including the spontaneous polarization and dielectric permittivity estimated from the GA optimized effective Landau coefficients are compared with the results obtained from first-principles calculations.
We also provide a qualitative comparison of the simulated domain dynamics during switching with available experimental observations~\cite{chouprik2018ferroelectricity}. Further, we introduce non-ferroelectric grains in the simulated grain structure to reduce the ferroelectric phase fraction and simulate the polarization behavior using the GA optimized effective Landau coefficients. The ferroelectric phase fraction is adjusted such that the simulated PE curve fits with the measured PE curve for a {Hf$_{0.75}$Zr$_{0.25}$O$_2$} thin film. We also use GA optimized effective Landau coefficients to investigate the effects of grain morphology and texture on switching characteristics of HZO thin films. 
\medskip    

\section{Experimental methods}
\medskip

\subsection*{Thin film fabrication}
\medskip

The $10~nm$-thick $Hf_{1-X}Zr_XO_2$ ($X = 0,0.25$ and $0.5$) films are deposited on the $90~nm$-thick TiN bottom electrode by atomic layer
deposition (Cambridge Nanotech Savannah S100) using tetrakis-dimethylamido-hafnium(IV), tetrakis-dimethylamido-zirconium(IV), and ozone ($O_3$) as the Hf precursor, Zr precursor, and oxygen source, respectively.
High concentration $O_3$ ($\sim 400~\frac{g}{m^3}$ ) used in this study was formed by an $O_3$ generator 
[OP-250H, Toshiba-Mitsubishi-Electric Industrial Systems Corporation]. 
The wafer temperature is set to $250~^\circ C$ during HZO deposition.
After deposition of $90~nm$-thick TiN top electrode at room temperature by radio frequency (RF) sputtering, the annealing process is performed for 60s at $300~^\circ C$, $400~^\circ C$, and $500~^\circ C$ in an $N_2$ atmosphere using a rapid thermal annealing system. Then,
{TiN/Hf$_{1-X}$Zr$_X$O$_2$/TiN} capacitors are
defined by conventional photolithography and etching processes performed using an Au hard mask [(Au ($85~nm$)/Pd ($3~nm$)] deposited by an electron-beam evaporator. The PE curves are measured at $10~kHz$ using a semiconductor parameter
analyzer (Keithley 4200-SCS). The average grain size is estimated from atomic force microscopy (AFM) images~\cite{kim2019stress}.
\medskip

\subsection*{Phase field model for polycrystalline HZO thin film}
\medskip

The proposed phase field model to simulate switching dynamics in polycrystalline HZO thin film describes the grain structure and the polar domain structure in every grain in the thin film~\cite{choudhury2005phase,choudhury2007effect,liu2013phase,su2015phase,vidyasagar2017predicting}. We generate the polycrystalline HZO thin film structure using a grain growth model based on the multi-phase field model proposed by Steinbach et al.~\cite{STEINBACH1996135,STEINBACH1999385}. The grain structure consisting of N grains is distinguished by grain order parameters $\phi_i(\textbf{r})$ assigned with a value of 1 in the $i'$th grain, 0 in other grains, and intermediate values at the grain boundaries. The free energy functional of the multi-grain system $F_\phi$ is expressed as a function of grain order parameters $\phi_i(\textbf{r})$ ($i=1,2,...,N$): 

\begin{equation}
    F_\phi=\int\sum_{i=1}^N \sum_{j=i+1}^N(W\phi_i\phi_j-\frac{a^2}{2}\nabla\phi_i \cdot\nabla\phi_j)dV,
\end{equation}

where $W$ is the height of the energy barrier and $a$ is the gradient coefficient of the interface between two grains. The temporal evolution of phase field variables are governed by the following set of equations,

\begin{equation}
    \frac{\partial\phi_i}{\partial t}=-\frac{2}{s}\sum_{j=1}^sM^\phi(\frac{\delta F_\phi}{\delta \phi_i}-\frac{\delta F_\phi}{\delta \phi_j}), \hspace{8mm}(i=1,2,...,N)
\end{equation}
\medskip

Here, $s$ is the number of active or non-zero order parameters at each grid point, and $M^\phi$  is the phase field mobility between two grains. Since we describe multiple grains with random orientations based on Euler angles~\cite{goldstein1953classical}, a common global coordinate system for every grain is defined. According to the rotation of the Euler angles, the transformation matrix from the global to local coordinate system $Tr_i$ can be expressed as follows: 

\begin{widetext}
\begin{equation}
Tr_i = 
\normalsize{\begin{bmatrix}
 \cos{\varphi_i}\cos{\psi_i}-\cos{\theta_i}\sin{\varphi_i}\sin{\psi_i} & \sin{\varphi_i}\cos{\psi_i}+\cos{\theta_i}\cos{\varphi_i}\sin{\psi_i} & \sin{\theta_i}\sin{\psi_i} \\
-\cos{\varphi_i}\sin{\psi_i}-\cos{\theta_i}\sin{\varphi_i}\cos{\psi_i} & -\sin{\varphi_i}\sin{\psi_i}+\cos{\theta_i}\cos{\varphi_i}\cos{\psi_i} & \sin{\theta_i}\cos{\psi_i} \\
\sin{\theta_i}\sin{\varphi_i} & -\sin{\theta_i}\cos{\varphi_i} & \cos{\theta_i}
\end{bmatrix}}
\end{equation}
\end{widetext}
\medskip

The polycrystalline structure generated using the multi-phase-field grain growth model does not evolve with time during polar domain evolution. The ferroelectric HZO thin films contain both polar and non-polar grains. For simplicity, we consider all the grains in the simulated microstructure to be ferroelectric. The domain structure within each grain is described as a function of the local polarization vector fields, $P_{Xi}(\textbf{r,t})$, $P_{Yi}(\textbf{r,t})$, and $P_{Zi}(\textbf{r,t})$. The total free energy of a ferroelectric polycrystal is given as follows:

\begin{equation}
    F= \int(f_{bulk}+f_{grad}+f_{ela}+f_{dep}+f_{appl} )dV,
\end{equation}

where $f_{bulk}$, $f_{grad}$, $f_{ela}$, $f_{dep}$, and $f_{appl}$ are the contributions from bulk free energy, gradient energy, elastic strain energy, depolarizing energy, and applied electric field, respectively. The bulk free energy density of the polycrystal is defined as follows:

\begin{equation}
    f_{bulk}=\sum_{i=1}^Nh(\phi_i)f_{Li}
\end{equation}

Here, $h(\phi_i)=\phi_i^3(10-15\phi_i+6\phi_i^2)$ and $f_{Li}$ is the bulk free energy density in each grain (identified by a subscript `$i$') expressed in terms of polarization components. We use Landau free energy for our ferroelectric model, assuming uniaxial directions of spontaneous polarization only in the out-of-plane direction because the polarization direction of the polar orthorhombic phase in HZO is along its c-axis~\cite{park2019modeling}.
Therefore, the polar and nonpolar contributions of the bulk free energy density functional are described using a one-dimensional Landau polynomial (expressed as a function of $P_{Zi}$) and a general quadratic form (expressed as a function of $P_{Xi}$ and $P_{Yi}$), respectively~\cite{saha2019phase,saha2020multi,park2019modeling}. Thus, the bulk energy density in a grain is given as follows:

\begin{equation}
    f_{L_i}=\alpha_1P_{Zi}^2+\alpha_{11}P_{Zi}^4+\alpha_{111}P_{Zi}^6+\frac{1}{2\chi_{f}}(P_{Xi}^2+P_{Yi}^2)
\end{equation}

Here, $\chi_f$ is the background isotropic dielectric susceptibility of the polar phase and $\alpha_1$, $\alpha_{11}$ and $\alpha_{111}$ are the Landau coefficients. The contributions from gradient energy $f_{grad}$ and applied electric field $f_{appl}$ are expressed as follows: 

\begin{equation}
\label{graden}
\begin{split}
     f_{grad}&=\frac{1}{2}G_{11}\sum_{i=1}^N\Big((P_{Xi,X})^2+(P_{Xi,Y})^2+(P_{Xi,Z})^2+(P_{Yi,X})^2\\
     &+(P_{Yi,Y})^2+(P_{Yi,Z})^2+(P_{Zi,X})^2+(P_{Zi,Y})^2+(P_{Zi,Z})^2\Big),
\end{split}
\end{equation}

\begin{equation}
\label{elecen}
    f_{appl}=-\sum_{i=1}^N(E_{Xi} P_{Xi}+E_{Yi} P_{Yi}+E_{Zi} P_{Zi}),
\end{equation}

where $G_{11}$ is the gradient energy coefficient, $P_{Ui,j}$ $(U=X,Y,Z)$ denotes the spatial derivative of $P_{Ui}$ with respect to the $j$th coordinate, and $E_{Ui}$ $(U=X,Y,Z)$ are the components of applied electric field along local coordinates. 
\medskip

The domain wall energy is dependent on the coefficients associated with the bulk free energy (Landau coefficients) and gradient energy ($G_{11}$) densities. Since spontaneous polarization is present only along uniaxial directions, $90^\circ $ domain walls are not considered in the model. The energy of $180^\circ$ domain walls is evaluated to be $\approx \frac{4}{3}P_0\sqrt{2G_{11}E_B}$, where $P_0$ is the spontaneous polarization ($P_0=\sqrt{\frac{-\alpha_{11}+\sqrt{\alpha_{11}^2-3\alpha_1\alpha_{111}}}{3\alpha_{111}}}$) and $E_B$ is the energy barrier ($E_B=\alpha_{11}P_0^4+2\alpha_{111}P_0^6$) for $180^\circ$ domain switching~\cite{elder2001sharp,hlinka2006phenomenological,marton2010domain}. The gradient energy coefficient $G_{11}$ is calculated from the energy of $180^\circ $ domain walls in HZO obtained from a first-principles investigation by Ding et al.~\cite{ding2020atomic}. Three types of $180^\circ $ domain walls were reported in their study. In our model, domain walls with negative energy and having mismatch along the Z-direction are not considered. We choose the $180^\circ $ domain wall without any lattice mismatch and having an energy of $0.2185~\frac{J}{m^{2}}$, which is the minimum value among reported energies.
\medskip

The elastic energy density in the polycrystal is given by

\begin{equation}
    f_{ela}=\frac{1}{2}\int \sigma_{ij}(\epsilon^T_{ij}-\epsilon^{0G}_{ij})dV,
\label{felast0}
\end{equation}

where $\sigma_{ij}$, $\epsilon^T_{ij}$, and $\epsilon^{0G}_{ij}$ denote the elastic stress, total strain, and spontaneous strain in global coordinate system, respectively. Subscripts 1, 2, and 3 denote Cartesian coordinates $X$, $Y$, $Z$ and Voigt's (matrix) notations are used. The spontaneous strain in a given grain can be expressed with respect to the local coordinate system in terms of electrostrictive tensor $Q_{ijkl}$. Since spontaneous polarization is considered only in the out-of-plane direction, spontaneous strain $\epsilon^0_i$ can be described in terms of $Q_{13}$, $Q_{23}$, and $Q_{33}$:

\begin{equation}
\epsilon^0_i = 
\begin{bmatrix}
Q_{13} & 0 & 0\\
0 & Q_{23} & 0\\
0 & 0 & Q_{33}\\
\end{bmatrix}
P_{Zi}^2
\end{equation}
\medskip

The spontaneous strain in the global coordinate system is related to the local spontaneous strain in terms of the transformation matrix by $\epsilon^{0G} = \sum_{i=1}^N\phi_iTr^\prime_i\epsilon^0_iTr_i$. The total strain is expressed as the sum of a macroscopic homogeneous strain $E_{ij}$ and a periodic homogeneous strain $\delta\epsilon_{ij}$:

\begin{equation}
\epsilon^T_{ij}=\delta\epsilon_{ij}+E_{ij}
\end{equation}
\medskip

We introduce a set of displacements to solve the heterogeneous strain such that $\delta\epsilon_{ij}=1/2(u_{i,j}+u_{j,i})$. The mechanical equilibrium condition given by $\sigma_{ij,j}=0$, is solved using the phase field microelasticity method~\cite{wang2002phase,wang2004phase,saj2019realization} (see Section S1 in the Supplementary Information). We do not solve elastic energy in the three dimensional GA calculations because of it's complexity and high computational cost. We consider the coupling terms between the strain and polarization to be included in the free energy coefficients for GA simulations. Under time dependent Ginzburg-Landau (TDGL) formalism, the polarization state equation for $P_{Zi}$ in a single grain can be expressed as~\cite{eliseev2012domain,saha2019phase}: 

\begin{equation}
\begin{split}
-\frac{1}{L}\frac{\partial P_{Zi}}{\partial t}=
&\frac{\partial \Big(\alpha_1P_{Zi}^2+\alpha_{11}P_{Zi}^4+\alpha_{111}P_{Zi}^6\Big)}{\partial P_{Zi}}\\
&-\frac{\partial \Big((Q_{13}\sigma_1+Q_{23}\sigma_2+Q_{33}\sigma_3)P^2_{Zi}\Big)}{\partial P_{Zi}}\\
&+\frac{\delta \Big(f_{grad}+f_{appl}\Big)}{\delta P_{Zi}},
\end{split}
\label{pzstate}
\end{equation}

where, $L$ is the kinetic coefficient related to domain wall mobility. Considering plane stress state $\sigma_3=0$,  we obtain the unknown stress components $\sigma_1$ and $\sigma_2$:

\begin{equation}
\begin{split}
&\sigma_1=\frac{(S_{22}Q_{13}-S_{12}Q_{23})(P^2_0-P^2_{Zi})}{S_{11}S_{22}-S_{12}^2}\\
&\sigma_2=\frac{(S_{11}Q_{23}-S_{12}Q_{13})(P^2_0-P^2_{Zi})}{S_{11}S_{22}-S_{12}^2}
\end{split}
\label{stress}
\end{equation}
\medskip

The derivation of eqn~(\ref{stress}) is provided in the Supplementary Information, Section S2. Substituting eqn~(\ref{stress}) in  the polarization state equation (eqn~(\ref{pzstate})) we obtain

\begin{equation}
\begin{split}
-\frac{1}{L}\frac{\partial P_{Zi}}{\partial t}=
&\frac{\partial \Big(\alpha_1P_{Zi}^2+\alpha_{11}P_{Zi}^4+\alpha_{111}P_{Zi}^6\Big)}{\partial P_{Zi}}\\
&-\frac{\Big(S_{11}Q_{23}^2+S_{22}Q_{13}^2-2S_{12}Q_{13}Q_{23}\Big)}{S_{11}S_{22}-S_{12}^2}\frac{\partial \Big((P^2_0-P^2_{Zi})P^2_{Zi}\Big)}{\partial P_{Zi}}\\
&+\frac{\delta \Big(f_{grad}+f_{appl}\Big)}{\delta P_{Zi}}
\end{split}
\label{pzstate2}
\end{equation}
\medskip

Eqn~(\ref{pzstate2}) can be rearranged by introducing $\eta=\frac{S_{11}Q_{23}^2+S_{22}Q_{13}^2-2S_{12}Q_{13}Q_{23}}{S_{11}S_{22}-S_{12}^2}$ to the form

\begin{equation}
\begin{split}
-\frac{1}{L}\frac{\partial P_{Zi}}{\partial t}=
&2\Big(\alpha_1-\eta P^2_0\Big)P_{Zi}+4\Big(\alpha_{11}+\eta\Big)P_{Zi}^3+6\alpha_{111}P_{Zi}^5\\
&+\frac{\delta \Big(f_{grad}+f_{appl}\Big)}{\delta P_{Zi}}
\end{split}
\label{pzstate3}
\end{equation}
\medskip

The polarization state equation can be rewritten in terms of effective Landau coefficients $a_1$, $a_{11}$, and $a_{111}$ as

\begin{equation}
-\frac{1}{L}\frac{\partial P_{Zi}}{\partial t}=2a_1P_{Zi}+4a_{11}P_{Zi}^3+6a_{111}P_{Zi}^5
+\frac{\delta \Big(f_{grad}+f_{appl}\Big)}{\delta P_{Zi}},
\label{pzstatef}
\end{equation}

where,

\begin{equation}
\begin{split}
a_1&=\alpha_1-\eta P_0^2\\
a_{11}&=\alpha_{11}+\eta\\
a_{111}&=\alpha_{111}
\label{effland}
\end{split}
\end{equation}

Experimental investigations have reported the presence of reduced depolarization fields in ferroelectric HZO thin films due to their ultra-thin thickness, low dielectric constant and large coercive field. This is evident from the symmetric and robust PE hysteresis curves in HZO thin films~\cite{gong2016fe,park2018review,zacharaki2020depletion}. Therefore, we do not consider depolarizing energy in our model for simplicity. The expression for the modified total free energy of a ferroelectric polycrystal described in terms of effective Landau coefficients is 

\begin{equation}
\label{toten}
F= \int(f_{bulk}+f_{grad}+f_{appl} )dV
\end{equation}

The simplified model reduces the computational complexity in comparison with the comprehensive model considering $90^\circ$ ferroelectric domains and additional elastic and depolarizing energy contributions. The spatio-temporal evolution of the domain structure can be obtained by solving the TDGL equations for local polarizations: 

\begin{equation}
\label{tdgl}
    \frac{\partial P_{Ui}}{\partial t}=-L\frac{\delta F}{\delta P_{Ui}}, \hspace{8mm}(U=X,Y,Z), \hspace{8mm}(i=1,2,...,N)
\end{equation}

The TDGL equations are discretized in space and time with finite difference and explicit forward Euler schemes, respectively. 
\medskip

To simulate switching dynamics, a domain structure is initially generated by giving random perturbations in the polarization field and evolved in the absence of an external electric field. Then, an external electric field $E^G_Z$ is applied along the global $Z$ direction at a rate matching the experimental sweeping rate to evolve the local polarization fields. For simulations, the electric field is incremented discretely after specific number of steps matching the frequency in experiments $10~kHz$. The magnitude of increment in applied electric field is $\Delta E=0.25~MV/cm$ and performed at an interval of 2500 number of steps which corresponds to a time interval of $2.5~\mu s$. The PE curve is generated by measuring the average global polarization along the direction of the applied electric field.
\medskip

The  polarizations and electric fields in the local coordinates are related to their respective counterparts ($P^G_{U},E^G_{U}; U=X,Y,Z$) in the global coordinate system through the transformation matrix: 

\begin{equation}
\begin{bmatrix}
E_{Xi} & E_{Yi} & E_{Zi}
\end{bmatrix}^\prime=
Tr_i\begin{bmatrix}
E^G_{X} & E^G_{Y} & E^G_{Z}
\end{bmatrix}^\prime
\end{equation}

\begin{equation}
\begin{bmatrix}
P_{Xi} & P_{Yi} & P_{Zi}
\end{bmatrix}^\prime=
Tr_i\begin{bmatrix}
P^G_{X} & P^G_{Y} & P^G_{Z}
\end{bmatrix}^\prime
\end{equation}
\medskip

We modify the phase field model to consider non-polar domains in the polycrystal to investigate the influence of the ferroelectric phase fraction on polarization switching. The bulk free energy density of the polycrystal with ferroelectric and non-ferroelectric grains can be expressed as follows: 

\begin{equation}
    f_{bulk}=\sum_{i=1}^n h(\phi_i)f_{Di}+\sum_{j=n+1}^N h(\phi_j)f_{Lj}
\end{equation}

Here, $N$ is the total number of grains in the thin film consisting of $n$ non-polar grains and the remaining polar grains. The bulk free energy density in a specific non-polar grain $f_{Di}$ can be expressed as follows:

\begin{equation}
    f_{D_i}=\frac{1}{2\chi_{d}}(P_{Xi}^2+P_{Yi}^2+P_{Zi}^2),
\end{equation}

where $\chi_{d}$ is the susceptibility of the non-ferroelectric phase, which is obtained by fitting the simulated polarization-electric field (PE) curve with the measured PE curve for a $10~nm$-thick {HfO$_2$} film.
\medskip

\subsection*{Genetic algorithm optimization}
\medskip

The GA is initiated with a population containing many sets of effective Landau coefficients and background dielectric susceptibilities, with each set representing a chromosome. The polarization switching curves are generated for each chromosome by phase field simulations. The objective function in our model measures the differences between the simulated polarization and experimental data~\cite{karr1995least}, which can be expressed as follows:

\begin{equation}    
    \Delta\rho=\sqrt{\frac{\sum_{i=0}^M\Big(\rho_s(e_i)-\rho_m(e_i)\Big)^2}{M}},
\end{equation}

where $\rho_s(e_i)$ and $\rho_m(e_i)$ are the simulated and measured polarizations, respectively, at an applied field $e_i$. $M$ is the number of external fields considered in the calculation of objective function. Since the coercive field and remnant polarization are the influential parameters defining the switching behaviour, the applied fields considered in calculating the objective function are limited by the following conditions: $0<e_i<-e_c$ and $e_c<e_i<0$. Here, $e_c$ is the measured coercive field for $10~nm$-thick {Hf$_{0.5}$Zr$_{0.5}$O$_2$} film.
\medskip

Chromosomes in every generation are classified according to the fitness of their objective functions and the two highest ranked chromosomes of a generation are selected as parents to reproduce the next generation through mutation and crossover~\cite{umbarkar2015crossover}. GA is stopped when the best chromosome does not change for many generations or an upper limit for the number of generations is reached. Thus, GA converges to a minimum for the objective function to predict effective Landau coefficients and background dielectric susceptibility for ferroelectric HZO.
\medskip

\subsection*{First-principles calculations}
\medskip

Our calculations are based on first-principles density functional theory (DFT) as implemented in the Vienna Ab-initio Simulation Package (VASP)~\cite{VASP}, with exchange correlation energy treated using the PBE functional ~\cite{pbe}. We use the Projector Augmented Wave (PAW) ~\cite{PAW} method to model the interaction between ionic cores and valence electrons and a plane wave basis for representing wavefunctions is truncated using a 500 eV cutoff. We use an 8x8x8 mesh of k-points in sampling the Brillouin zone integrations. The optimized lattice parameters are a=5.04 \AA{}, b= 5.078 \AA{} \& c= 5.26 \AA{} for the ferroelectric phase, in agreement with previous DFT calculations and experiments~\cite{kunneth2019thermodynamics}. Berry phase method and density functional perturbation theory is used to calculate the spontaneous polarization and dielectric permittivity of ferroelectric phase of hafnia~\cite{vaspoptic}. The elastic constants are calculated using the finite differences method~\cite{vaspelastic}.
\medskip

\section{Results and discussion}
\medskip

The three-dimensional phase field simulations are performed using $\Delta X\times \Delta Y\times t_Z$ box with discrete grids, $\Delta x=\Delta y=\Delta z=1.0~nm$ at a time increment $\Delta t=1~ns$. We assume periodic boundary conditions along $X$ and $Y$ directions and Dirichlet boundary condition along the $Z$ direction with polarizations considered to be zero outside the boundaries. The parameters for simulating switching dynamics with calibrated (Set I) and GA optimized (Set II) effective Landau coefficients are listed in Table \ref{set1} and Table \ref{set2}, respectively. Table \ref{set34} provides the simulation parameters used to investigate the effect of grain morphology (Set III) on polarization switching. For simulations to investigate the influence of texture, the parameters are set to be the same as those in Set II, but $[001]$ and $[111]$ fiber textured grains are considered instead of randomly oriented grains. In the case of simulations to study the effect of ferroelectric phase fraction, only the fractions of polar grains are set to be less than unity ($\nu_0=0.38,0.5,$ and $0.8$) and the other simulation parameters are the same as those in Set II.
\medskip

\begin{table}[h]
\small
\centering
   \caption{Simulation parameters used to predict switching dynamics with calibrated (Set I) Landau polynomials}
  \begin{tabular*}{0.48\textwidth}{@{\extracolsep{\fill}}ll}
    \hline
    Parameter &  Value \\
    \hline
    Thin film surface area ($\Delta X\times\Delta Y$)  &  $640\times640~nm^2$  \\
    Thin film thickness ($t_Z$)  &  $10~nm$  \\
    Effective Landau coefficient ($a_1$)  &  $-2.976\times 10^8~\frac{Jm}{C^2}$  \\
    Effective Landau coefficient ($a_{11}$)  &  $-2.160\times 10^8~\frac{Jm^5}{C^4}$  \\
    Effective Landau coefficient ($a_{111}$)  &  $1.653\times 10^{10}~\frac{Jm^9}{C^6}$  \\
    Background dielectric susceptibility ($\chi_f$)  &  $2.568\times 10^{-10}~\frac{C^2}{Jm}$  \\
    Gradient energy coefficient ($G_{11}$)  &  $9.788\times 10^{-9}~\frac{Jm^3}{C^2}$  \\
    Kinetic coefficient ($L$)  &  $9.326\times 10^{-2}~\frac{C^2}{Jms}$  \\
    Average grain size ($D$)  &  $30~nm$  \\
    Number of grains ($N$)  &  $472$  \\
    \hline
  \end{tabular*}
\label{set1}
\end{table}

\begin{table}[h]
\small
\centering
   \caption{Simulation parameters used to predict switching dynamics with GA optimized (Set II) Landau polynomials}
  \begin{tabular*}{0.48\textwidth}{@{\extracolsep{\fill}}ll}
    \hline
    Parameter & Value   \\
    \hline
    Thin film surface area ($\Delta X\times\Delta Y$) &  $640\times640~nm^2$  \\
    Thin film thickness ($t_Z$)  & $10~nm$  \\
    Effective Landau coefficient ($a_1$)  & $-4.289\times 10^8~\frac{Jm}{C^2}$ \\
    Effective Landau coefficient ($a_{11}$)  & $-2.242\times 10^8~\frac{Jm^5}{C^4}$  \\
    Effective Landau coefficient ($a_{111}$)  & $2.170\times 10^9~\frac{Jm^9}{C^6}$ \\
    Background dielectric susceptibility ($\chi_f$)   & $4.019\times 10^{-10}~\frac{C^2}{Jm}$ \\
    Gradient energy coefficient ($G_{11}$)  & $5.066\times 10^{-10}~\frac{Jm^3}{C^2}$ \\
    Kinetic coefficient ($L$)  & $9.326\times 10^{-2}~\frac{C^2}{Jms}$ \\
    Average grain size ($D$)  & $30~nm$ \\
    Number of grains ($N$)  & $472$ \\
    \hline
  \end{tabular*}
\label{set2}
\end{table}

\begin{table}[h]
\small
\centering
   \caption{Simulation parameters used to investigate the effect of grain morphology (Set III) on polarization switching}
  \begin{tabular*}{0.48\textwidth}{@{\extracolsep{\fill}}ll}
    \hline
    Parameter &  Value\\
    \hline
    Thin film surface area ($\Delta X\times\Delta Y$)  &  $480\times480~nm^2$\\
    Thin film thickness ($t_Z$)  &  $20~nm$\\
    Effective Landau coefficient ($a_1$)  &  $-4.289\times 10^8~\frac{Jm}{C^2}$\\
    Effective Landau coefficient ($a_{11}$)  &  $-2.242\times 10^8~\frac{Jm^5}{C^4}$ \\
    Effective Landau coefficient ($a_{111}$)  &  $2.170\times 10^9~\frac{Jm^9}{C^6}$ \\
    Background dielectric susceptibility ($\chi_f$)  &  $4.019\times 10^{-10}~\frac{C^2}{Jm}$ \\
    Gradient energy coefficient ($G_{11}$)  &  $5.066\times 10^{-10}~\frac{Jm^3}{C^2}$\\
    Kinetic coefficient ($L$)  &  $9.326\times 10^{-2}~\frac{C^2}{Jms}$ \\    
    Average grain size ($D$)  &  $20~nm$ \\
    Number of grains ($N$)  &  $555,427$\\
    \hline
  \end{tabular*}
\label{set34}
\end{table}

Additionally, we perform phase field simulations considering elastic energy for polycrystalline ferroelectrics. The Landau coefficients ($\alpha_1$, $\alpha_{11}$, and $\alpha_{111}$) estimated using eqn~(\ref{effland}) corresponding to GA optimized effective Landau coefficients ($a_1$, $a_{11}$, and $a_{111}$) are used for these simulations. The elastic constants of ferroelectric HZO are computed from first-principles calculations. The values of the elastic constants and electrostriction coefficients used for the phase field simulations are: $C_{11}=400.92~GPa$, $C_{12}=127.10~GPa$, $C_{13}=133.00~GPa$, $C_{22}=397.30~GPa$, $C_{23}=99.30~GPa$, $C_{33}=353.30~GPa$, $C_{44}=91.66~GPa$, $C_{55}=84.17~GPa$, $C_{66}=127.67~GPa$, $Q_{13}=-0.02~\frac{m^4}{C^2}$, $Q_{23}=-0.015~\frac{m^4}{C^2}$, and $Q_{33}=0.030~\frac{m^4}{C^2}$. Note that, in this work, 
due to lack of proper experimental/first-principles based statistics, 
we assume the electrostrictive coefficients in a similar manner described by Glinchuk et al.~\cite{glinchuk2020origin}
\medskip

\subsection*{Phase field simulation using calibrated effective Landau coefficients}
\medskip

\begin{figure}[h]
\centering
  \includegraphics[width=0.6\linewidth]{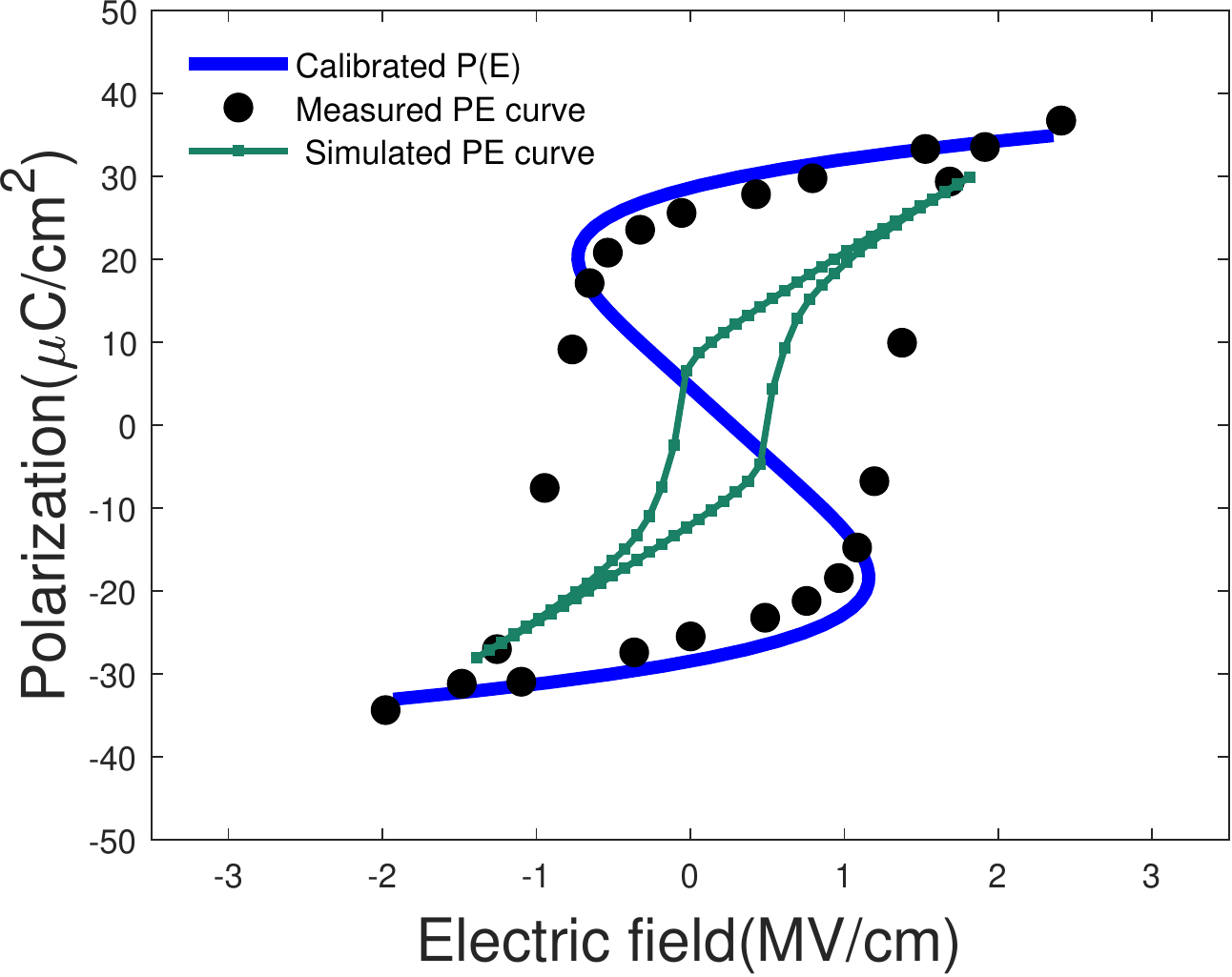}
  \caption{Switching characteristics simulated using calibrated effective Landau coefficients. Comparison of calibrated $P(E)$ and the simulated PE curve with the measured PE curve for $10~nm$-thick {Hf$_{0.5}$Zr$_{0.5}$O$_2$} film. Measured PE curve reprinted (adapted) with permission from (Kim, Si Joon, et al. \textit{ACS applied materials $\&$ interfaces} 11.5 (2019): 5208-5214)~\cite{kim2019stress}. Copyright (2019) American Chemical Society.}
  \label{lksim}
\end{figure}

We calibrate polarization hysteresis function with measured PE data for a $10~nm$-thick {Hf$_{0.5}$Zr$_{0.5}$O$_2$} film~\cite{kim2019stress} with average grain size, $D$ of $\sim 30~nm$ and estimate the effective Landau coefficients. The measured values of remnant polarization and coercive field are $\sim 25.65~\frac{\mu C}{cm^2}$ and $\sim 1.1~\frac{M V}{cm}$, respectively. 
The polarization switching in ferroelectrics can be described by the TDGL equation:
\begin{equation}
\label{lk}
    \frac{\partial P}{\partial t}=-\Gamma\frac{\delta F_P}{\delta P},
\end{equation}
where $P$ is the domain polarization, $t$ is time, $\Gamma$ is the kinetic coefficient, and $F_P$ is the total energy of the ferroelectric system. The total energy can be expressed as function of polarization, applied electric field ($E$) and effective Landau coefficients:
\begin{equation}
    F_P=a_1P^2+a_{11}P^4+a_{111}P^6-EP
\end{equation}
The inverse of $P(E)$ is derived by assuming static equilibrium ($\frac{\partial P}{\partial t}=0$) and substituting total energy into the TDGL equation (eqn~(\ref{lk})):
\begin{equation}
    E=2a_1P+4a_{11}P^3+6a_{111}P^5
\end{equation}
The intrinsic polarization hysteresis function is a 'S' shaped curve. It can be fitted with the measured PE curve generated during polarization switching to extract the effective Landau coefficients. The calibrated effective Landau coefficients are $a_1=-2.976\times 10^8~\frac{Jm}{C^2}$, $a_{11}=-2.160\times 10^8~\frac{Jm^5}{C^4}$, and $a_{111}=1.653\times 10^{10}~\frac{Jm^9}{C^6}$ (see Table S1 and Fig. S1 in the Supplementary Information). The calibrated polarization hysteresis function and the measured PE curve are plotted in Fig.~\ref{lksim}.  
\medskip

\begin{figure*}[h]
  \centering
\begin{subfigure}[t]{.35\textwidth}
 \caption{}
  \centering
  \includegraphics[height=4.7cm]{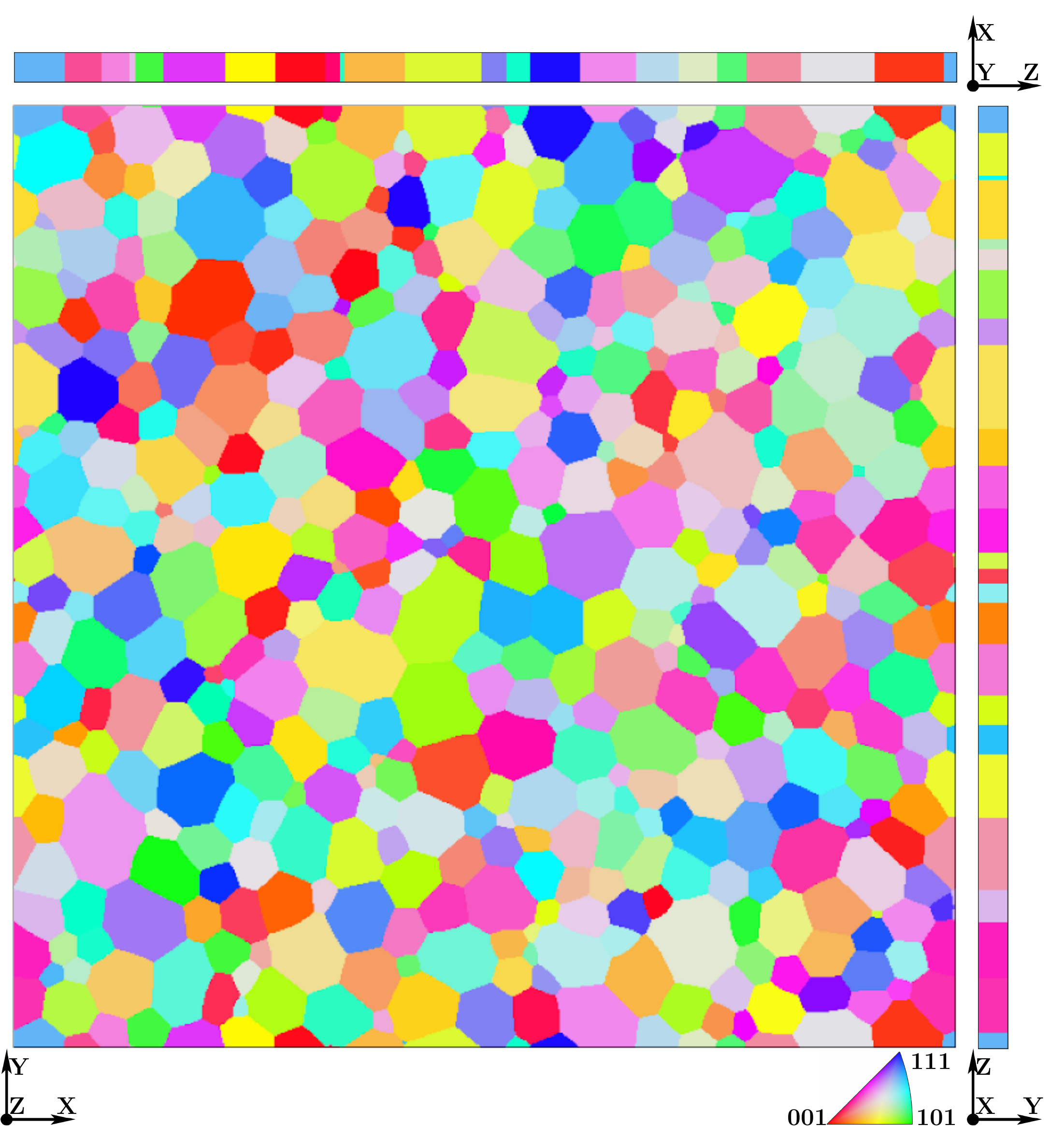}
   \label{fig:ga_1}
\end{subfigure}
  \centering
\begin{subfigure}[t]{.35\textwidth}
\caption{}
  \centering
  \includegraphics[height=4.7cm]{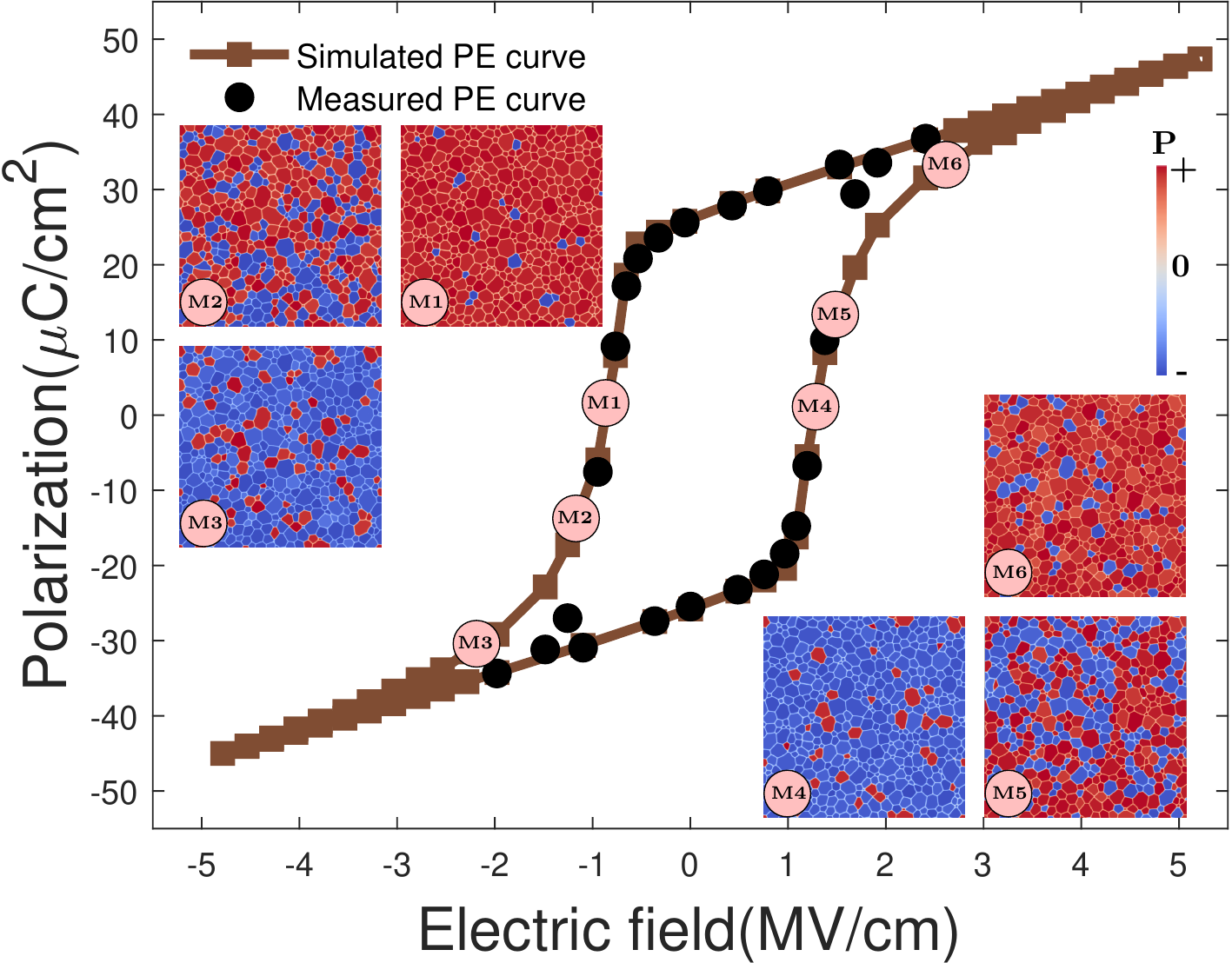}
   \label{fig:ga_2}
\end{subfigure}
\caption{Switching characteristics simulated using GA optimized effective Landau coefficients. (a) Top view ($XY$ plane) and side views ($YZ$ and $ZX$ planes) of simulated columnar thin film microstructure ($640\times640\times10~nm^3, D=30~nm$). (b) Comparison of simulated PE curve and experimentally measured curve for $10~nm$-thick {Hf$_{0.5}$Zr$_{0.5}$O$_2$} film. Domain structures at specific points along the PE loop as indicated by M1-M6 (inset). Measured PE curve (Panel \textit{b}) reprinted (adapted) with permission from (Kim, Si Joon, et al. \textit{ACS applied materials $\&$ interfaces} 11.5 (2019): 5208-5214)~\cite{kim2019stress}. Copyright (2019) American Chemical Society.}
\label{fig:ga}
\end{figure*}

Switching dynamics is simulated using calibrated effective Landau coefficients in polycrystalline thin film. We generate a thin film with specifications and grain morphology matching the experimental specimen and containing randomly oriented grains. Fig.~\ref{fig:ga}(a) shows the simulated columnar thin film microstructure. Background dielectric susceptibility ($\chi_f=2.568\times 10^{-10}~\frac{C^2}{Jm}$) is calculated from the dielectric constant ($\kappa\sim30$) of the polar phase in ferroelectric HZO taken from the literature~\cite{hyuk2013evolution}. The relation $\chi_f=\varepsilon_0(\kappa-1)$ is employed for computation of background dielectric susceptibility, where $\varepsilon_0=8.854\times 10^{-12}~\frac{C^2}{Jm}$ is the vacuum permittivity. The simulated PE curve given in Fig.~\ref{lksim} shows significant deviations from the experimental data with considerably low coercive field ($\sim 0.3~\frac{M V}{cm}$) and remnant polarization ($\sim 10.15~\frac{\mu C}{cm^2}$). The presence of grain boundaries substantially reduces the polarization and shrinks the PE hysteresis loop in polycrystalline ferroelectrics. 
\medskip

\subsection*{Phase field simulations coupled with genetic algorithm to optimize effective Landau coefficients}
\medskip

\begin{figure}[h]
\centering
  \includegraphics[width=0.6\linewidth]{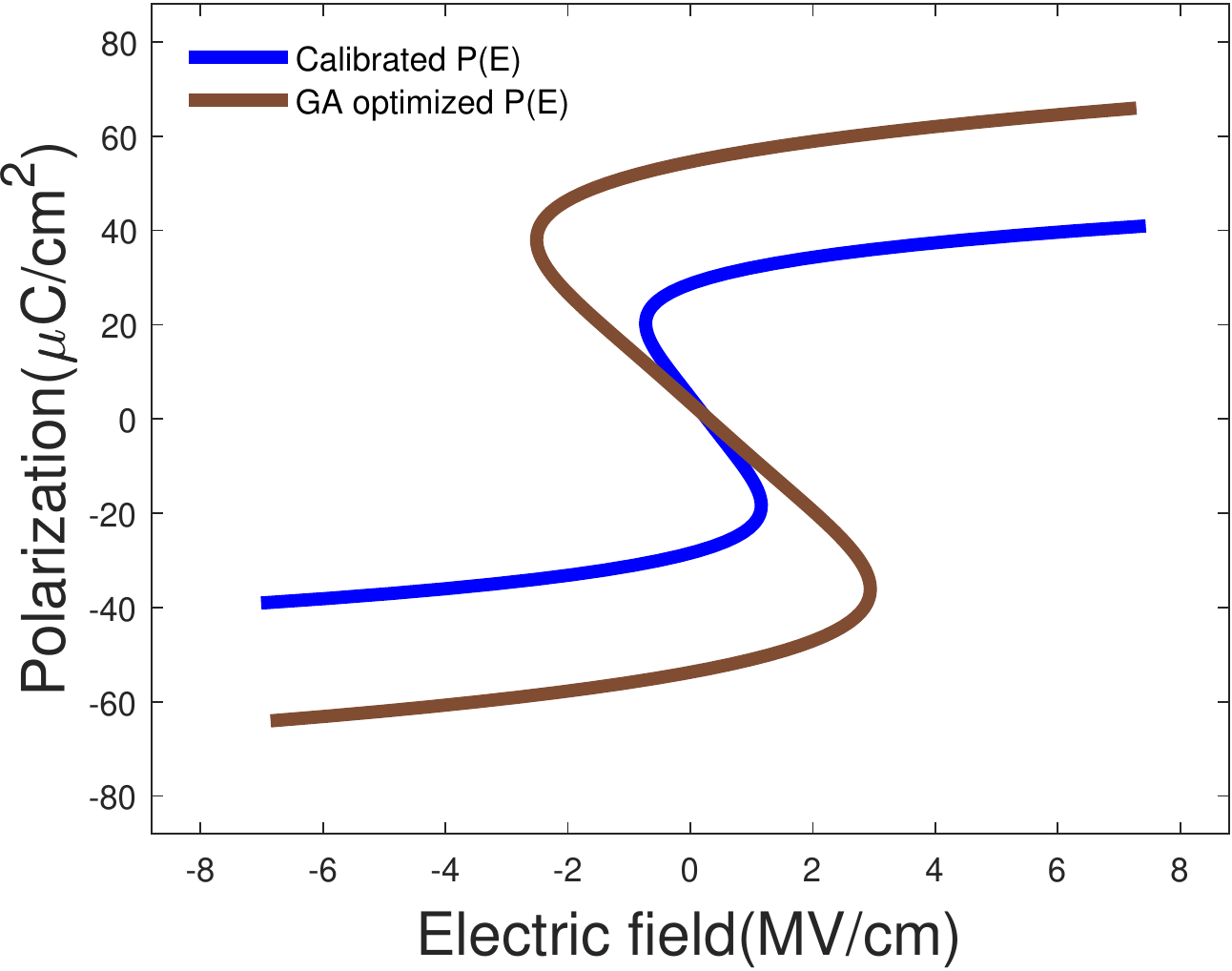}
  \caption{Comparison of polarization hysteresis functions. The calibrated $P(E)$ is compared with $P(E)$ for GA optimized effective Landau polynomials.}
  \label{comp_lk}
\end{figure}

The polycrystalline ferroelectric phase field model when coupled with the GA optimizes the Landau polynomial by minimizing inconsistencies between the simulated and measured switching curves. The GA converges to an optimized set of effective Landau coefficients and background dielectric susceptibility. The GA optimized effective Landau coefficients and dielectric susceptibility are $a_1=-4.289\times 10^8~\frac{Jm}{C^2}$, $a_{11}=-2.242\times 10^8~\frac{Jm^5}{C^4}$, $a_{111}=2.170\times 10^9~\frac{Jm^9}{C^6}$, and $\chi_f=4.019\times 10^{-10}~\frac{C^2}{Jm}$ (see Table S2 and Fig. S2 in the Supplementary Information). The simulated curve fits much better with the measured data as illustrated in Fig.~\ref{fig:ga}(b) demonstrating similar values of coercive field ($\sim 1.1~\frac{M V}{cm}$) and remnant polarization ($\sim 25.67~\frac{\mu C}{cm^2}$). However, some discrepancies are observed in the saturation region. This can be attributed to intrinsic defects in the ferroelectrics, which are not included in our model. 
Fig.~\ref{comp_lk} demonstrates the comparison between polarization hysteresis functions for calibrated and GA optimized effective Landau coefficients.
The optimized polarization hysteresis function has a higher switching voltage and much larger polarizations under the same applied electric fields than the calibrated polarization hysteresis function.
\medskip

The single-domain ground state property of HZO is estimated from the GA optimized effective Landau coefficients (see the Supplementary Information, Section S5). The values of spontaneous polarization and relative dielectric permittivity are $54.16~\frac{\mu C}{cm^2}$ and $43.2$, respectively. The spontaneous polarization and dielectric permittivity acquired from first-principles calculations are $51~\frac{\mu C}{cm^2}$ and $25.8$, respectively. The value of spontaneous polarization calculated from effective Landau coefficients is in good agreement with the first-principles calculation results. The deviation in the value of dielectric permittivity may be due to the assumption of uniaxial direction of spontaneous polarization in the phase field model. Even though we obtained the effective Landau coefficients considering the thin film to be $100\%$ ferroelectric, the fraction of polar orthorhombic phase in HZO is observed to be lower in experiments~\cite{kim2020comprehensive,mukundan2021ferroelectric}. This also may be a reason for the deviation in dielectric permittivity values.
\medskip

\begin{figure*}[h]
\centering
  \includegraphics[width=0.75\linewidth]{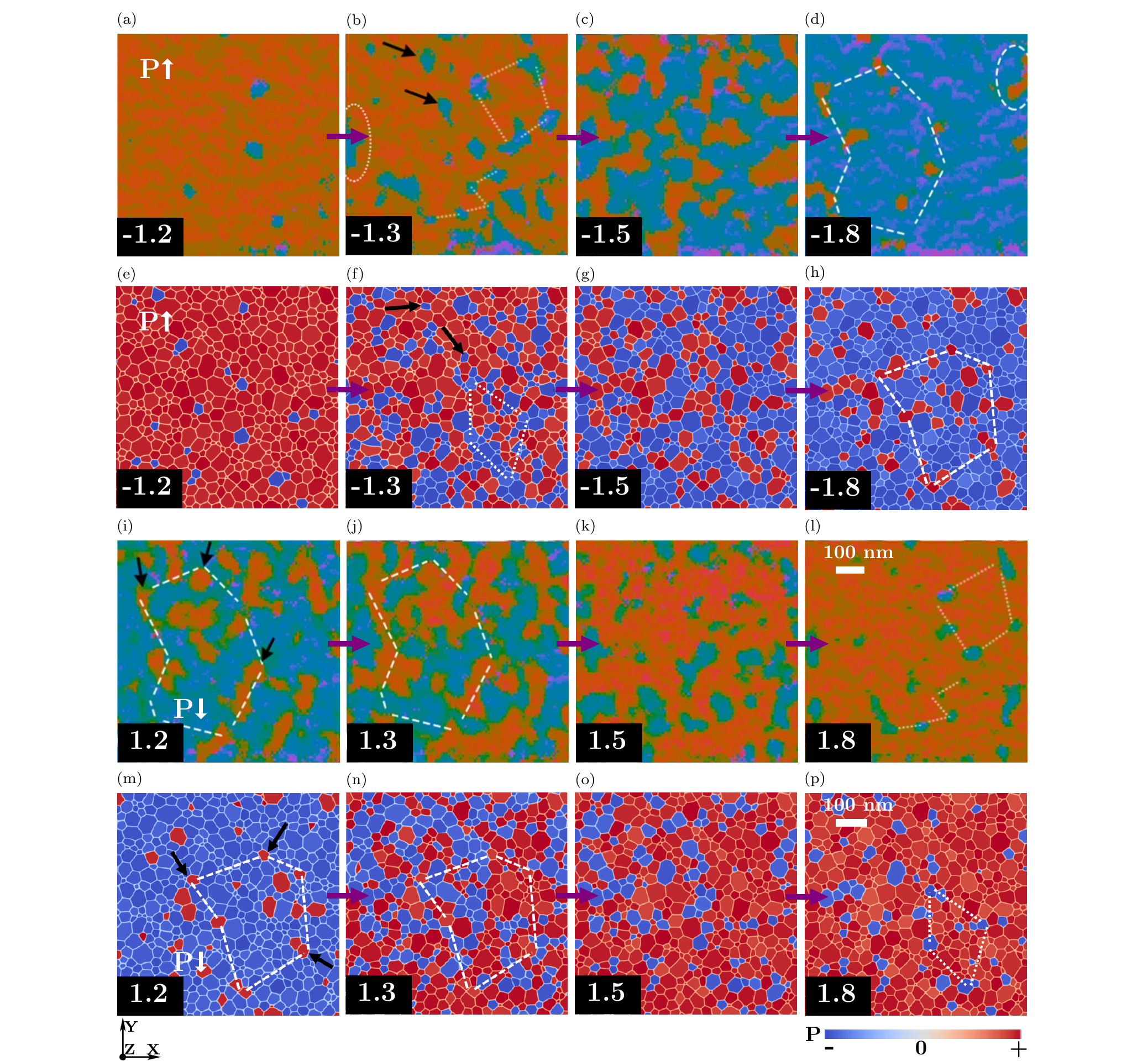}
\caption{Comparison of domain dynamics during switching in experiment and simulation (violet arrows indicate the sequence in which electric field is applied during switching). (a-d) Normalized BE PFM/AFAM phase maps obtained from experiments and (e-h) corresponding polarization profiles generated from phase field simulations for the ferroelectric thin film (top surface) during polarization reversal (up $\to$ down) under negative applied electric fields. (i-l) Normalized BE PFM/AFAM phase maps and (m-p) corresponding polarization profiles during polarization reversal (down $\to$ up) under positive applied electric bias. Black arrows indicate the location of domain seeds with the opposite direction of the polarization vector; white dashed and dotted lines indicate the localization of domains with preferred up and down polarization, respectively. The domain structures from experiments and simulations are compared at equivalent values of normalized applied electric field (inset). Panels \textit{a-d}, and \textit{i-l} reprinted (adapted) with permission from (Chouprik, Anastasia, et al. \textit{ACS applied materials $\&$ interfaces} 10.10 (2018): 8818-8826)~\cite{chouprik2018ferroelectricity}. Copyright (2018) American Chemical Society.}
\label{fig:domevol}
\end{figure*}

We qualitatively compare the domain dynamics of ferroelectric HZO thin film predicted by simulations with the available experimental observation by Chouprik et al~\cite{chouprik2018ferroelectricity}. They investigated the switching phenomenon in $10~nm$-thick {Hf$_{0.5}$Zr$_{0.5}$O$_2$} film and analysed the domain structures by resonance-enhanced combined band-excitation piezoresponse force microscopy (BE PFM) and atomic force acoustic microscopy (BE AFAM) techniques. Further, they normalized PFM data on AFAM data to generate the normalized BE PFM/AFAM phase maps of the thin film during ferroelectric switching. The comparison between the domain dynamics obtained from literature and simulated polarization profiles are shown in Fig.~\ref{fig:domevol}. For better understanding, we compare the domain structures from both experiments and simulations at applied electric fields normalized with respect to corresponding switching fields (see Table S3 in the Supplementary Information). During switching under negative applied bias, the polarization reversal from up to down is observed in both experimental ({Fig.~\ref{fig:domevol}(a)-(d)}) and simulated ({Fig.~\ref{fig:domevol}(e)-(h)}) domain morphology. Similarly, the polarization reverses from down to up during switching under positive electrical loading in the domain structures obtained from experiments ({Fig.~\ref{fig:domevol}(i)-(l)}) and phase field simulations ({Fig.~\ref{fig:domevol}(m)-(p)}). The domain dynamics during switching predicted by phase field modeling agree well with the experimental observations. It is evident from both simulated and experimental results that the reversal of the polarization in the ferroelectric thin film occurs by the nucleation and growth of the opposite polarization domains. 
\medskip

\begin{figure}[h]
\centering
  \includegraphics[width=0.6\linewidth]{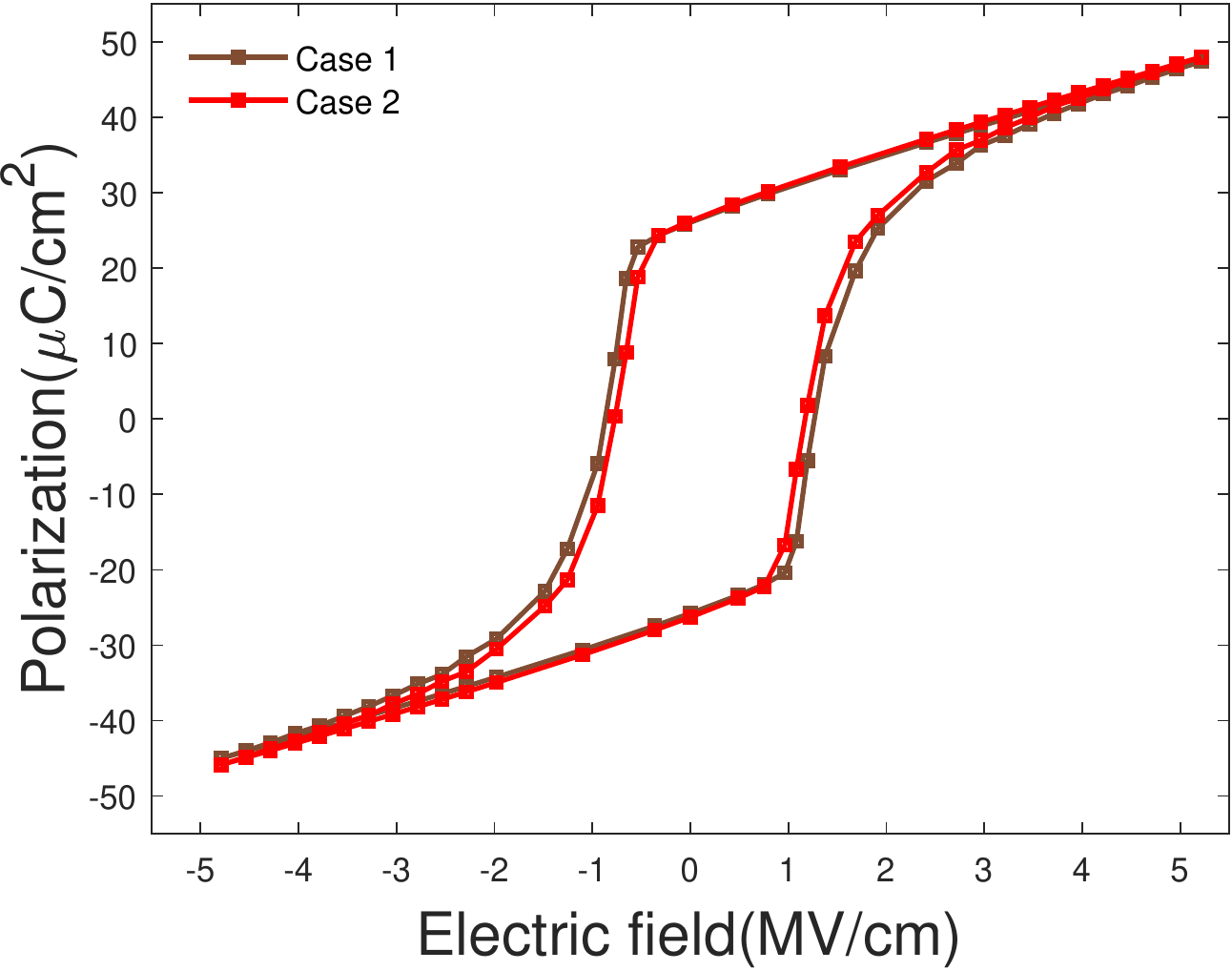}
    \caption{Comparison of PE curves for polycrystalline thin film generated using GA optimized effective Landau coefficients in the absence of elastic interactions (Case 1) and corresponding Landau coefficients considering elastic interactions (Case 2).}
  \label{polcomp}
\end{figure}

Further, we simulate the switching behaviour considering elastic interactions in polycrystalline thin films using Landau coefficients calculated from GA predicted effective Landau coefficients (eqn~(\ref{effland})). The Landau coefficients used for the switching simulations are: $\alpha_1=-3.475\times 10^8~\frac{Jm}{C^2}$, $\alpha_{11}=-5.015\times 10^8~\frac{Jm^5}{C^4}$, and $\alpha_{111}=1.653\times 10^{10}~\frac{Jm^9}{C^6}$. The switching curves generated using effective Landau coefficients in the absence of elastic interactions (Case 1) and Landau coefficients considering elastic energy (Case 2) are compared in Fig.~\ref{polcomp}. The PE curve generated for the effective Landau coefficients matches with the curve for corresponding Landau coefficients except for some discrepancies in the coercive field. This shows that the switching behaviour predicted using the simplified model does not deviate much from the comprehensive model considering Landau coefficients incorporating elastic energy contributions. Moreover, the values of steady state polarizations across $180^\circ$ domain walls and switching curves in a bulk single crystal are also matching for the effective Landau coefficients and corresponding Landau coefficients (see Fig. S3 and S4 in the Supplementary Information).
\medskip

\begin{figure}[h]
\centering
  \includegraphics[width=0.6\linewidth]{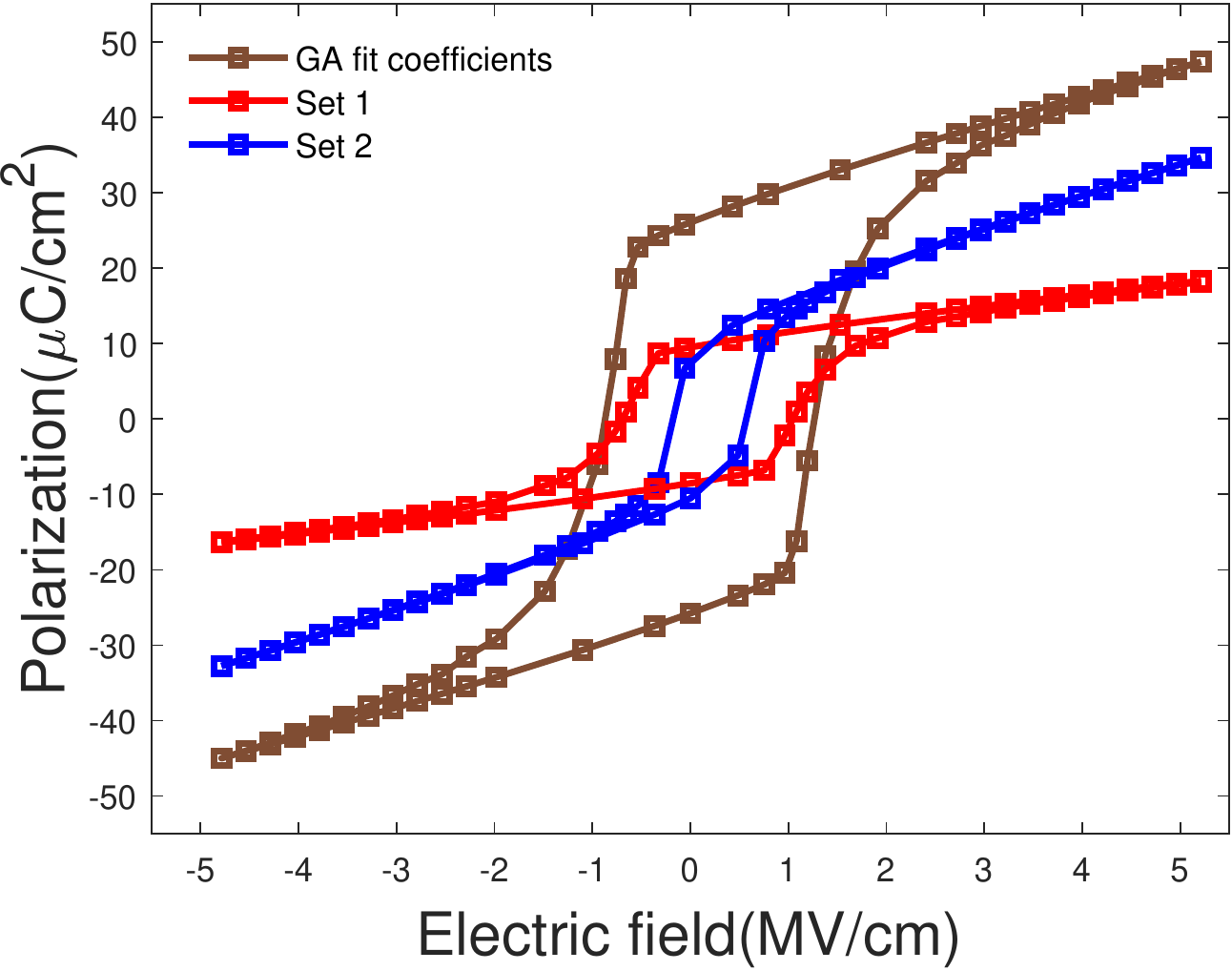}
  \caption{Comparison of PE curves for polycrystalline thin films simulated using GA optimized effective Landau coefficients and Landau coefficients reported in existing literature (Set 1 and 2).}
  \label{comp_lit}
\end{figure}

We simulate polarization switching in polycrystalline thin films using Landau coefficients reported in existing literature and compare with the switching curves generated using GA optimized effective Landau coefficients. Two sets of Landau coefficients obtained from previous studies are used in the simulations: $a_1=-1.25\times 10^9~\frac{Jm}{C^2}$, $a_{11}=1.5\times 10^{10}~\frac{Jm^5}{C^4}$, $a_{111}=2.5\times 10^{10}~\frac{Jm^9}{C^6}$ (Set 1~\cite{saha2020multi}) and $a_1=-4.0\times 10^8~\frac{Jm}{C^2}$, $a_{11}=3.7\times 10^9~\frac{Jm^5}{C^4}$, $a_{111}=1.1\times 10^9~\frac{Jm^9}{C^6}$ (Set 2~\cite{hsu2020theoretical}). The PE curves simulated using Landau coefficients obtained from literature shows significant deviations from the PE curve generated using GA optimized effective Landau coefficients (see Fig.~\ref{comp_lit}) with considerably low coercive fields and remnant polarizations. Landau coefficients reported in literature are calibrated coefficients obtained by fitting polarization hysteresis function with the measured PE curve. In polycrystalline thin film, the PE hysteresis loop shrinks due to the presence of grain boundaries. Therefore, the calibrated Landau coefficients exhibit lower coercive field and remnant polarization compared with GA optimized effective Landau coefficients. 
\medskip

\subsection*{Effect of ferroelectric phase fraction on the switching characteristics}
\medskip

\begin{figure}[h]
  \centering
\begin{subfigure}{.3\textwidth}
  \caption{}
  \centering
  \includegraphics[width=1.0\linewidth]{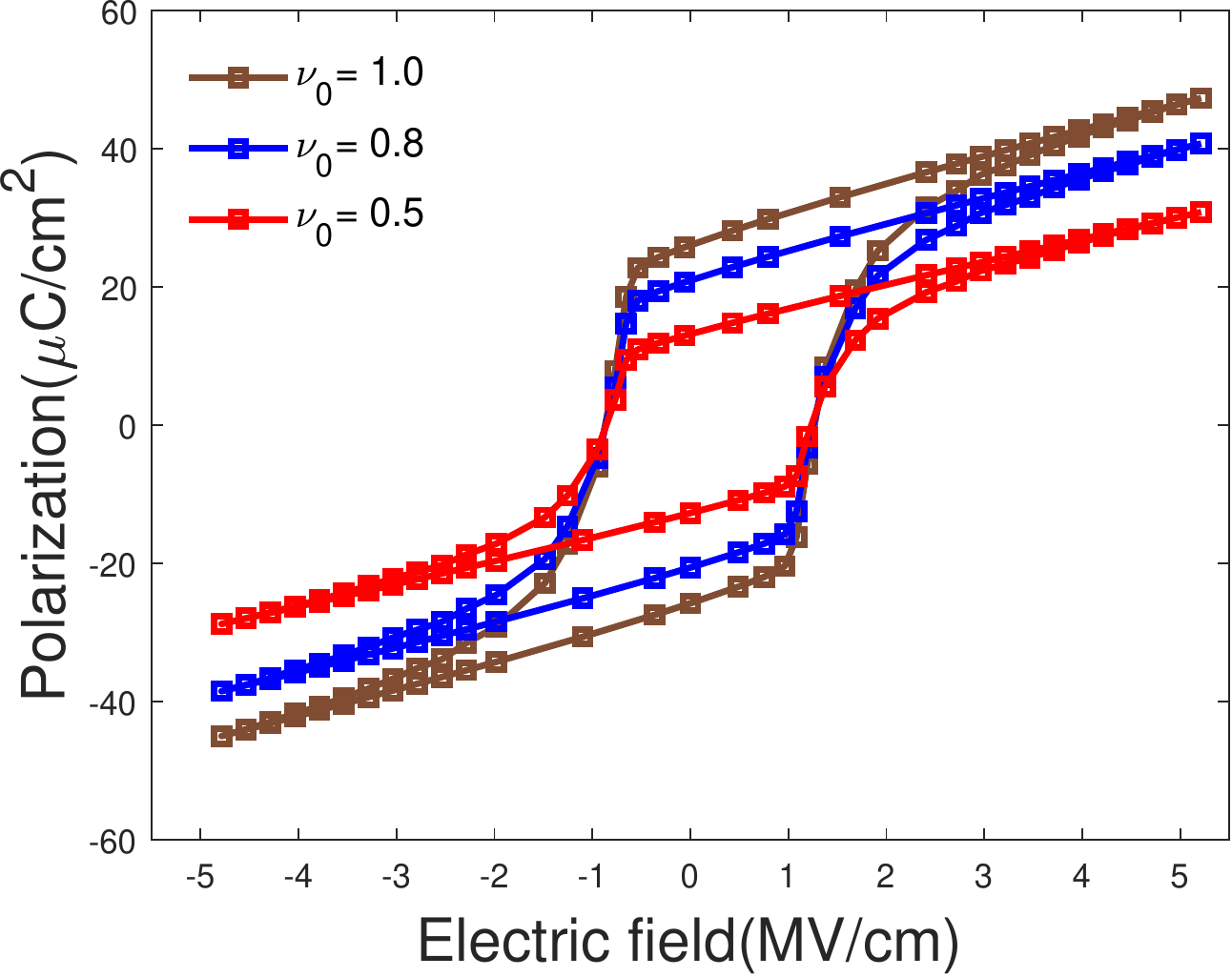}
   \label{fig:fefrac_1}
\end{subfigure}
\begin{subfigure}{.3\textwidth}
  \caption{}
  \centering
  \includegraphics[width=1.0\linewidth]{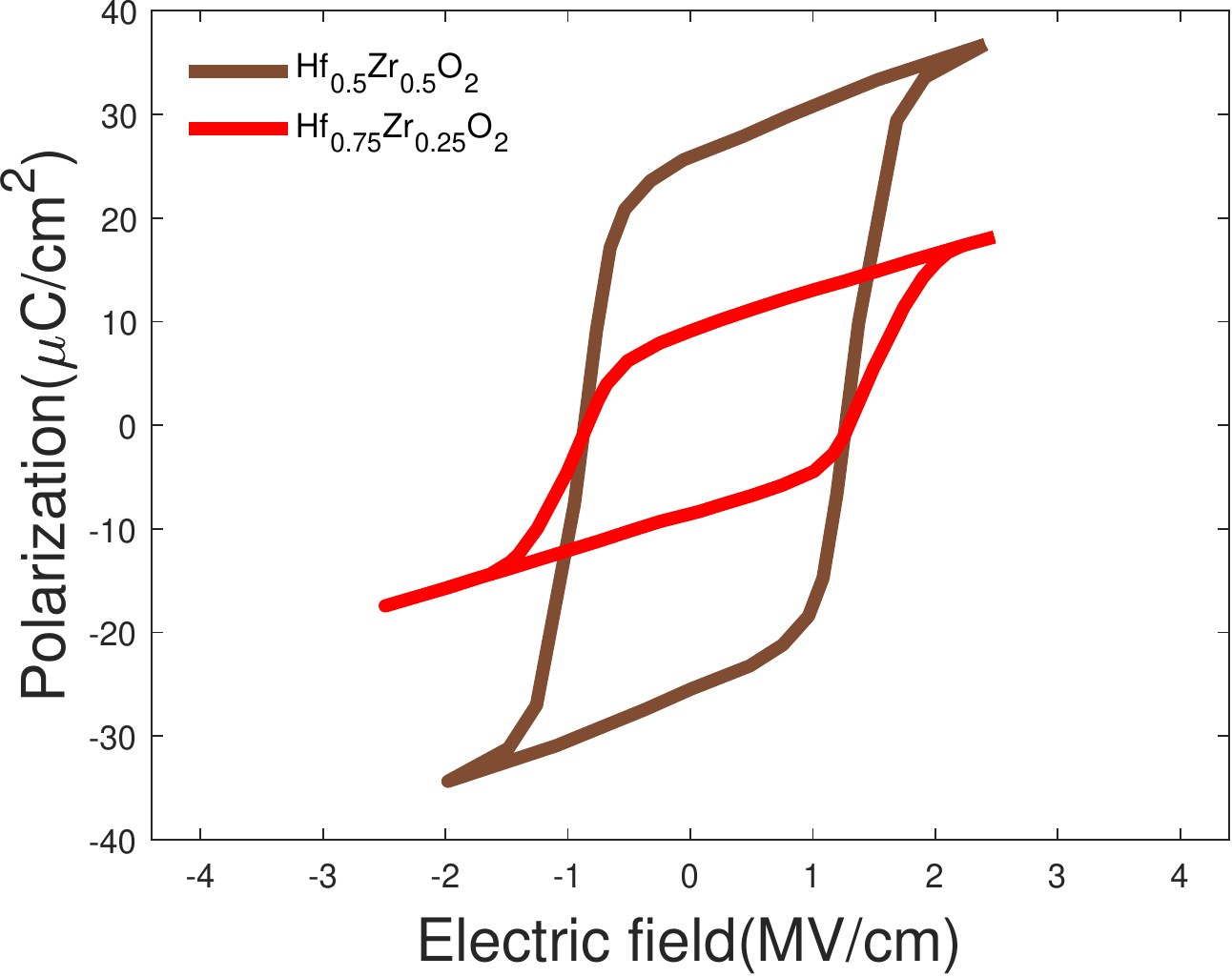}
   \label{fig:fefrac_2}
\end{subfigure}
\caption{Effect of polar phase fraction on switching characteristics. (a) Comparison of simulated PE curves for columnar thin films with different ferroelectric phase fractions ($\nu_o=0.5, 0.8,$ and $1.0$). (b) Comparison of measured PE curves for HZO thin films with different Zr concentrations. Panel \textit{b} reprinted (adapted) with permission from (Kim, Si Joon, et al. \textit{ACS applied materials $\&$ interfaces} 11.5 (2019): 5208-5214)~\cite{kim2019stress}. Copyright (2019) American Chemical Society.}
\label{fig:fefrac}
\end{figure}

To reduce the ferroelectric phase fraction in the polycrystalline structure, we introduce non-ferroelectric grains and perform simulations using a modified phase field model, which also considers non-ferroelectric domains. The dielectric susceptibility of non-polar grains ($\chi_d=2.108\times 10^{-10}~\frac{C^2}{Jm}$) is extracted from the measured PE curve of $10~nm$-thick {HfO$_2$} film (Fig. S5 in the Supplementary Information). We vary the fraction of ferroelectric grains ($\nu_0=0.5$ and $0.8$) and simulate polarization switching. The resulting PE curves are plotted in Fig.~\ref{fig:fefrac}(a).
The decrease in the ferroelectric phase ratio reduces remnant polarization without significantly affecting the coercive field. When the polar phase fraction varies from $0.8$ to $0.5$, remnant polarization decreases from $\sim 20.75~\frac{\mu C}{cm^2}$ to $\sim 12.91~\frac{\mu C}{cm^2}$, but the coercive field remains unchanged ($\sim 1.1~\frac{M V}{cm}$). We correlate these simulation results with measured PE curves for $10~nm$-thick {Hf$_{0.5}$Zr$_{0.5}$O$_2$} and {Hf$_{0.75}$Zr$_{0.25}$O$_2$} films. From Fig.~\ref{fig:fefrac}(b), it is evident that a decrease in the Zr concentration reduces remnant polarization, but does not change the coercive field. The values of remnant polarization and coercive field for {Hf$_{0.75}$Zr$_{0.25}$O$_2$} thin film are $\sim 8.90~\frac{\mu C}{cm^2}$ and $\sim 1.1~\frac{M V}{cm}$, respectively. 

\begin{figure}[h]
\centering
  \includegraphics[width=0.6\linewidth]{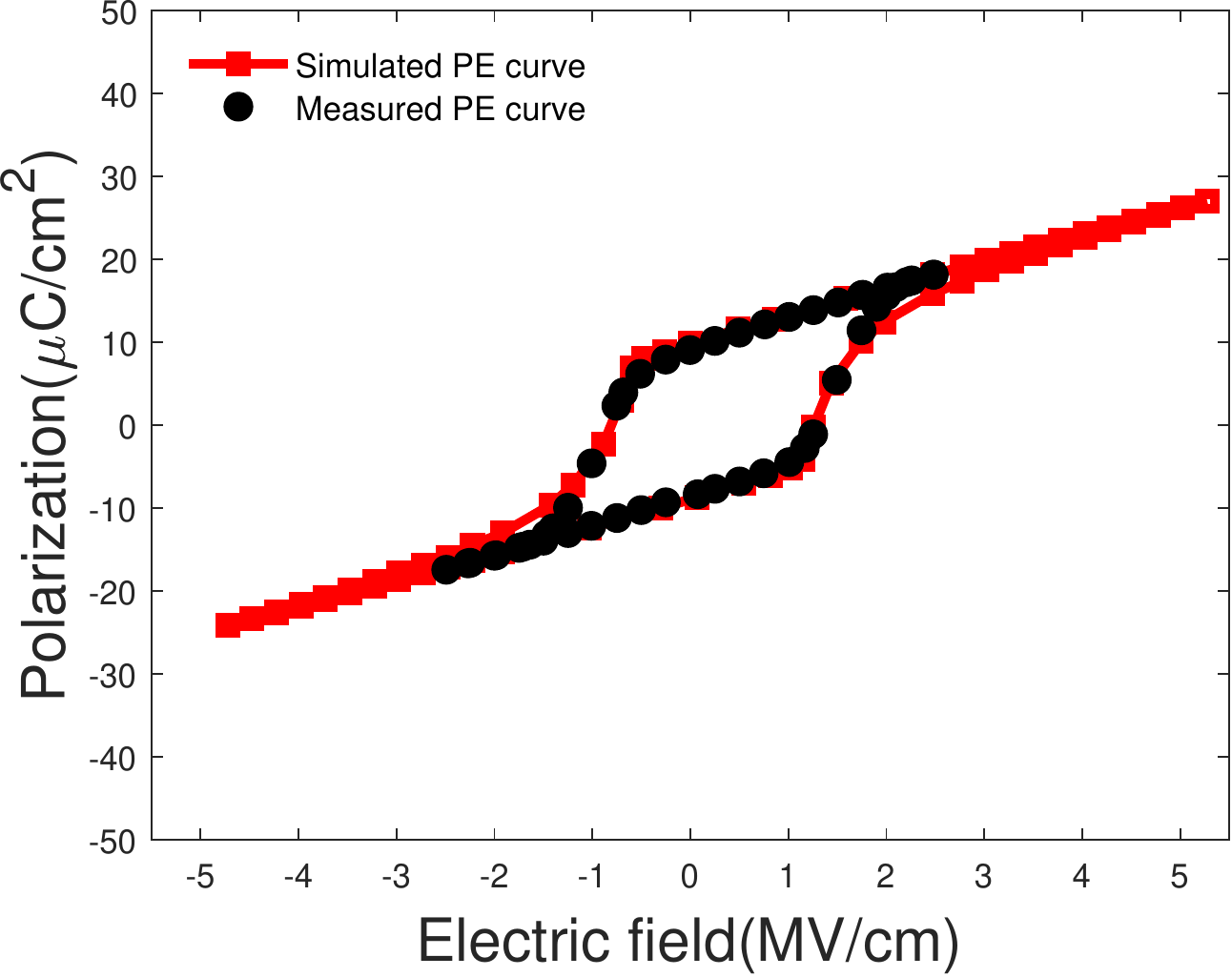}
  \caption{Switching characteristics for columnar thin film with $38\%$ ferroelectric grains. Comparison of simulated PE
  curve with the measured PE curve for $10~nm$-thick {Hf$_{0.75}$Zr$_{0.25}$O$_2$} film. Measured PE curve reprinted
  (adapted) with permission from (Kim, Si Joon, et al. \textit{ACS applied materials $\&$ interfaces} 11.5 (2019):
  5208-5214)~\cite{kim2019stress}. Copyright (2019) American Chemical Society.}
  \label{fedi_fit}
\end{figure}

We vary the fraction of ferroelectric grains and estimate the ferroelectric phase ratio at which the simulated PE curve matches the measured hysteresis loop for $10~nm$-thick {Hf$_{0.75}$Zr$_{0.25}$O$_2$} film. The simulated curve for a thin film with $38\%$ ferroelectric grains fits well with the measured curve exhibiting similar values of coercive field ($\sim 1.1~\frac{M V}{cm}$) and remnant polarization ($\sim 8.81~\frac{\mu C}{cm^2}$). The comparison between the simulated and measured curves are shown in Fig.~\ref{fedi_fit}. When we change the ferroelectric phase fractions, the simulated PE curves generated using the same effective Landau coefficients match with the measured PE curves for HZO thin films having different Zr concentrations. This indicates that the switching behavior in HZO thin films can be simulated efficiently by a simplified phase field model considering only $180^\circ$ ferroelectric domains with bulk free energy described by effective Landau coefficients. The model shows good performance at a low computational cost. There is plenty of scope to improve the model by extending it to a comprehensive model considering $90^\circ$ ferroelectric domains and additional elastic and depolarizing energy contributions.
\medskip

The value of $\chi_d$ used in simulations can be expressed as a weighted average of dielectric susceptibilities of non-polar phases ($1.682\times 10^{-10}~\frac{C^2}{Jm}$ and $3.010\times 10^{-10}~\frac{C^2}{Jm}$ for monoclinic and tetragonal phases, respectively) estimated from the dielectric constants ($\sim19$ and $\sim35$ for monoclinic and tetragonal phases, respectively) found in the literature~\cite{hyuk2013evolution}. This implies that the ratio between non-polar phases in HZO thin films may not vary with changes in Zr concentration. The simulations are consistent with the general belief that remnant polarization can be increased by more ferroelectric phase formation to enhance ferroelectricity in HZO films~\cite{goh2018enhanced,persson2020reduced}. However, contrary to our assumption that {Hf$_{0.5}$Zr$_{0.5}$O$_2$} thin films contain $100\%$ ferroelectric grains, the fraction of polar orthorhombic phase reported in experiments is much lower~\cite{park2017surface,schroeder2019recent,kim2020comprehensive,mukundan2021ferroelectric}. So, there is plenty of scope for improving the ferroelectricity in HZO thin films by maximizing the ferroelectric phase fraction. Further investigations on stabilizing the polar orthorhombic phase in HZO thin films are required. 
\medskip

\subsection*{Effect of crystalline texture and grain morphology on the switching characteristics}
\medskip

\begin{figure}[h]
\centering
  \includegraphics[width=0.65\linewidth]{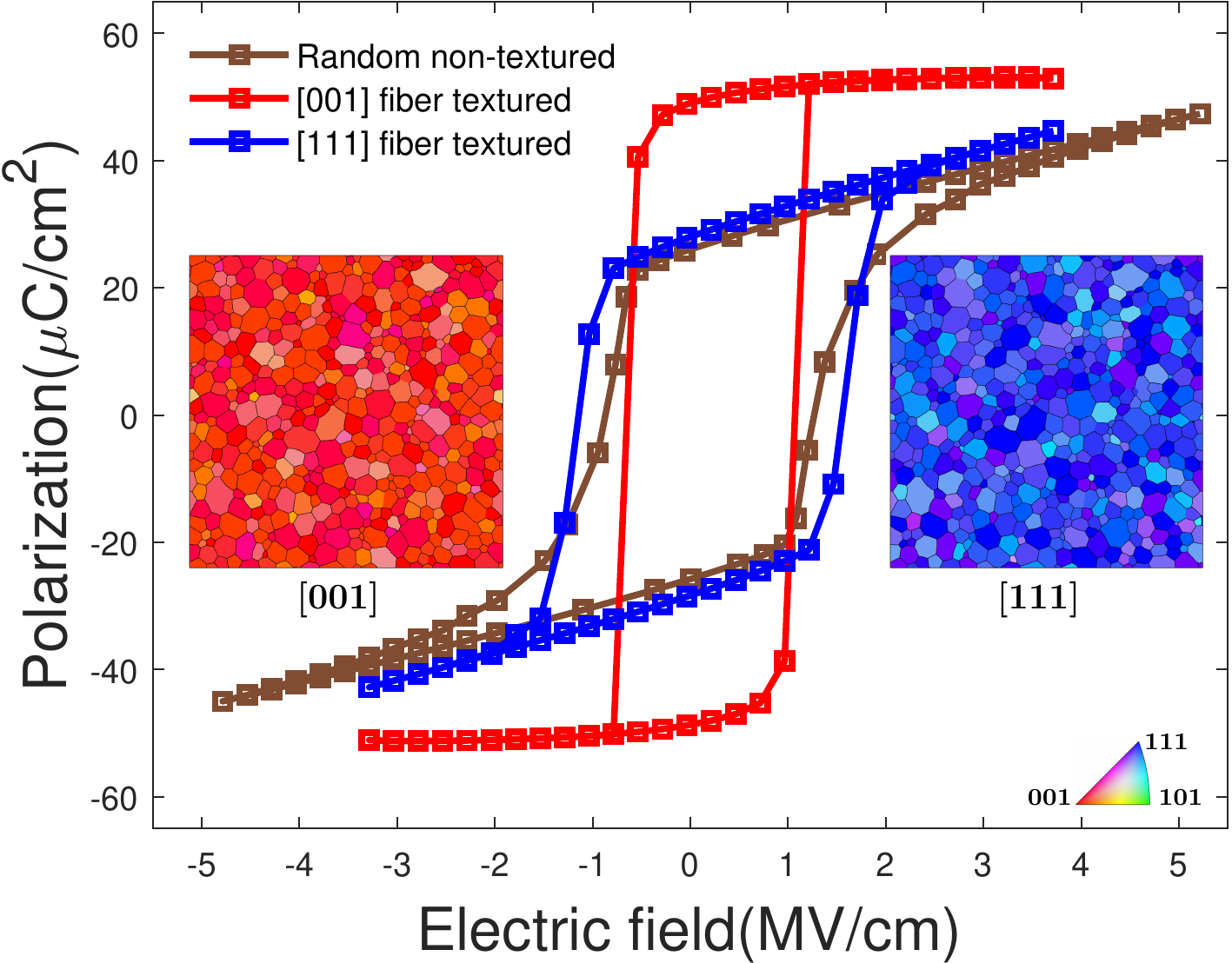}
\caption{Effect of crystalline texture on switching characteristics. Comparison of simulated PE curves for textured thin films with the PE curve of randomly oriented thin film. Orientation maps (top view) of textured thin films (inset).}
  \label{texture_col}
\end{figure}

Polarization switching is simulated in $10~nm$-thick columnar films considering $[001]$ and $[111]$ fiber textured grains instead of randomly oriented grains to understand the influence of crystalline texture on coercive field and remnant polarization. Fig.~\ref{texture_col} shows the comparison of PE curves for textured and non-textured thin films. The $[001]$ fiber textured thin film has lower coercive field ($\sim 0.9~\frac{M V}{cm}$) and higher remnant polarization ($\sim 48.94~\frac{\mu C}{cm^2}$) values than its non-textured counterpart. Whereas, the estimated values ($E_c\sim 1.3~\frac{M V}{cm}$ and $P_r\sim 28.20~\frac{\mu C}{cm^2}$) are higher for $[111]$ fiber textured thin film than the random non-textured thin film. Additionally, the PE curve for [001] textured film is compared with the PE curve for single crystalline HZO (see the Supplementary Information, Fig. S6).
\medskip

In textured thin films, the shapes of switching curves become bilinear. These films also exhibit fast switching transition and large remnant polarization because the crystallographic axes of textured grains do not deviate from the direction of the external electric field like the axes of randomly oriented grains in a non-textured thin film. Since the direction of the external electric field is along the direction of spontaneous polarization in $[001]$ fiber textured thin film, it has the largest remnant polarization. This indicates that control of crystallographic texture is highly beneficial for enhancement of ferroelectricity in HZO thin films. Moreover, the large coercive field, which is a limitation in many applications of ferroelectric HZO, can also be reduced by crystalline texture engineering.
\medskip

\begin{figure*}[h]
  \centering
\begin{subfigure}[t]{.35\textwidth}
\caption{}
  \centering
  \includegraphics[height=4.7cm]{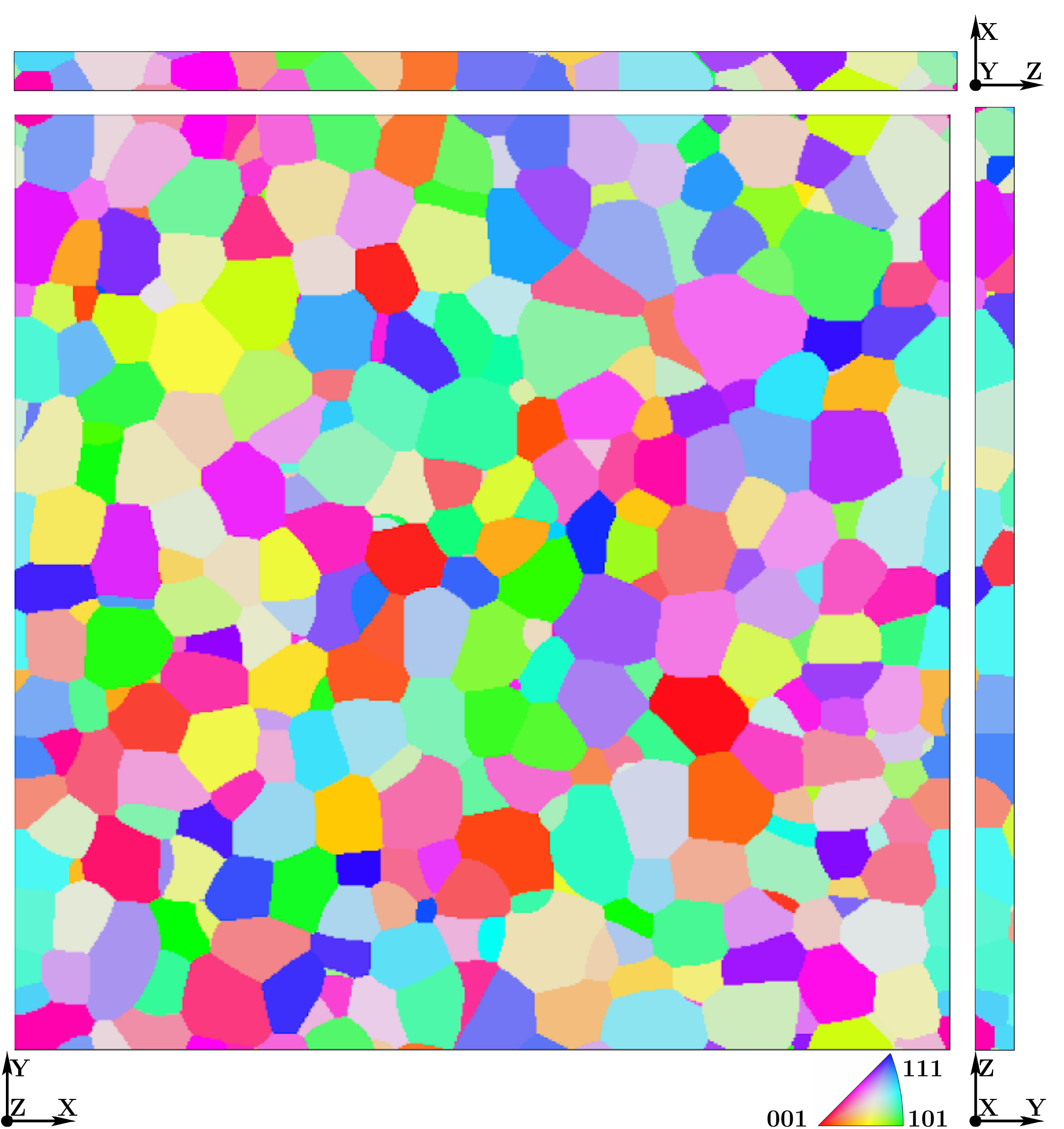}
   \label{morph_1}
\end{subfigure}
\begin{subfigure}[t]{.35\textwidth}
\caption{}
  \centering
  \includegraphics[height=4.7cm]{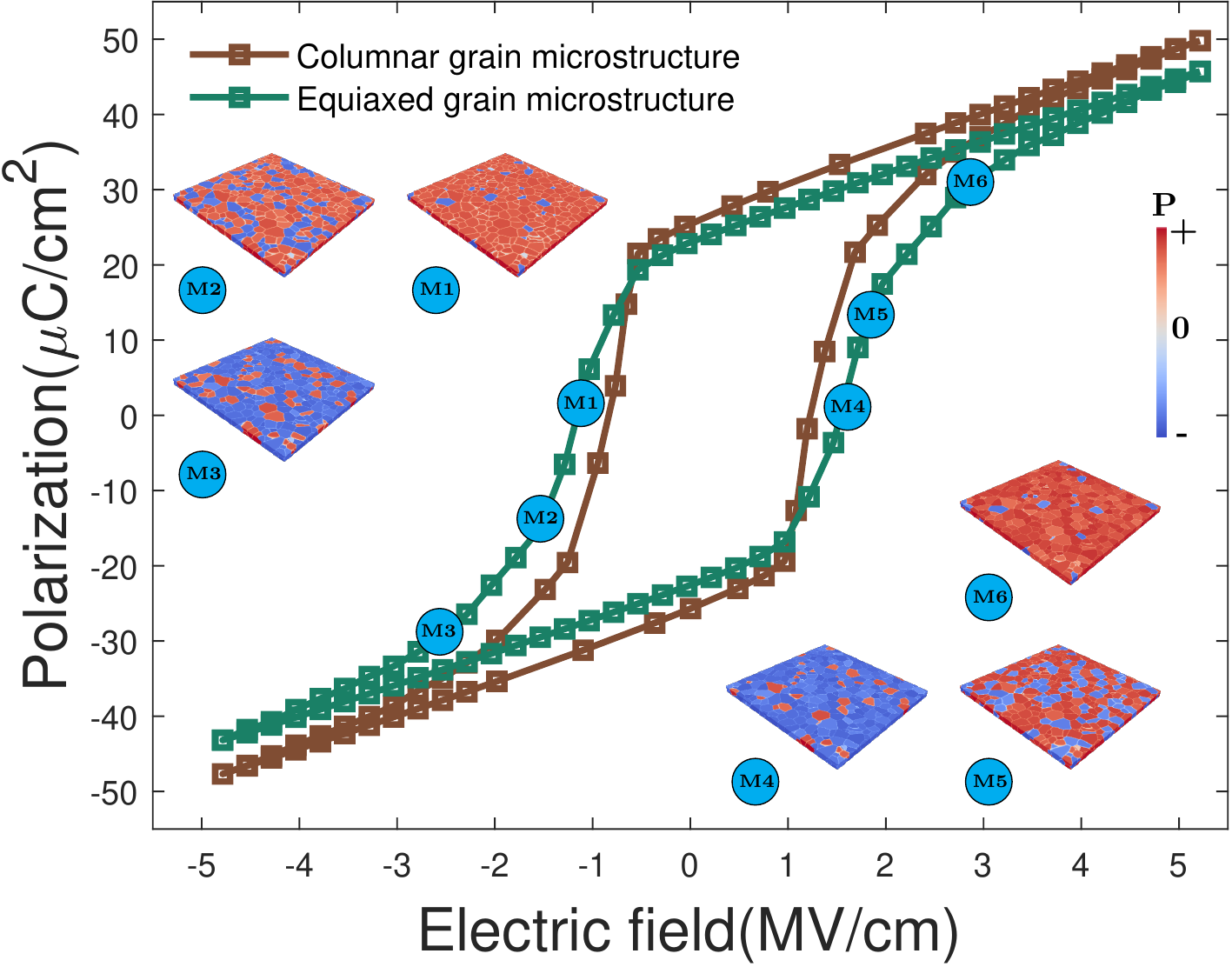}
   \label{morph_2}
\end{subfigure}
  \caption{Effect of grain morphology on switching characteristics. (a) Top view ($XY$ plane) and side views ($YZ$ and $ZX$ planes) of simulated microstructure of equiaxed HZO thin film ($480\times480\times20~nm^3, D=20~nm$). (b) Comparison of simulated PE curves for thin films with columnar and equiaxed grain morphology. Domain structures for the equiaxed thin film at specific points along the PE loop as indicated by M1-M6 (inset).}
\label{fig:morph}
\end{figure*}

Simulations are performed on $20~nm$-thick columnar and equiaxed (see Fig.~\ref{fig:morph}(a) for simulated equiaxed thin film microstructure) films with a grain size of $20~nm$ to investigate the influence of grain morphology on switching phenomena. Indeed, the switching dynamics change with grain morphology as illustrated in Fig.~\ref{fig:morph}(b) and thin film with equiaxed structure has reduced remnant polarization ($\sim 22.80~\frac{\mu C}{cm^2}$), but increased coercive field ($\sim 1.3~\frac{M V}{cm}$) compared with columnar thin film ($P_r\sim 25.29~\frac{\mu C}{cm^2}$, $E_c\sim 1.1~\frac{M V}{cm}$).
\medskip

The lower remnant polarization in equiaxed thin films compared to thin films with columnar grain microstructure can be attributed to the presence of grain boundaries across the direction of the applied electric field. Whereas, the grain boundaries are found only along the direction of the applied electric field in columnar thin films. The absence of grain boundaries across the the direction of applied field results in large remnant polarization and fast switching in thin films with columnar grains. The columnar grain morphology with increased remnant polarization and reduced coercive field enhances ferroelectricity and removes constraints due to the large coercive field in HZO thin films. Therefore, in HZO ferroelectrics, columnar thin films would be preferred over thin films with equiaxed grain microstructure. Moreover, the effect of grain size on polarization switching in thin films with equiaxed and columnar grain morphology is also investigated (see Fig. S7 and S8 in the Supplementary Information).
\medskip

\section{Conclusions}
\medskip

To summarize, by combining a simplified polycrystalline phase field model with GA, an efficient method is developed to estimate the effective Landau coefficients for ferroelectric HZO thin films. The discrepancies between the simulated and measured data observed in previous computational models describing polarization switching in HZO are rectified in our model. The simulated PE curve generated using GA optimized effective Landau coefficients shows excellent agreement to the experimental PE curve for the values of the coercive field and remnant polarization. The nucleation and growth of opposite polarization domains leading to the reversal of polarization during switching observed in phase field simulations are consistent with available experimental findings~\cite{chouprik2018ferroelectricity}. Moreover, the phase field model is validated by simulating polarization switching in a thin film with a lower ferroelectric phase fraction using optimized effective Landau coefficients and matching the simulated PE curve with the measured PE curve for an HZO thin film with a lower Zr concentration. The simulations also demonstrate that ferroelectricity can be enhanced and the coercive field related constraints can be eliminated in HZO thin films by control of crystallographic texture. Further, the simulations indicate that columnar grain morphology in HZO thin films is preferable to equiaxed grain morphology for ferroelectric applications. The present work can be extended to predict effective Landau coefficients and simulate polarization switching in other hafnia based ferroelectric thin films.   
\medskip

\section*{Acknowledgements}

This research was supported by National Research Foundation of Korea (NRF) grant funded by Ministry of Science and ICT (MSIT) of the Republic of Korea (Nos. NRF-2019R1A2C1089593, NRF-2020M3H4A3106736, NRF-2021M3H4A6A01045764).



\balance


\bibliography{rsc} 
\bibliographystyle{rsc} 
\clearpage


\section*{Supplementary material}

# Supplementary Information for

# A phase field model combined with genetic algorithm for polycrystalline hafnium zirconium oxide ferroelectrics

*Sandeep Sugathan [a], Krishnamohan Thekkepat [b,c], Soumya Bandyopadhyay [a], Jiyoung Kim [d] and Pil-Ryung Cha [a],\**

[a] School of Advanced Materials Engineering, Kookmin University, Seoul 02707, Korea.

[b] Electronic Materials Research Center, Korea Institute of Science and Technology, Seoul 02792, Republic of Korea.

[c] Division of Nano & Information Technology, KIST School, Korea University of Science and Technology, Seoul 02792, Republic of Korea.

[d] Department of Materials Science and Engineering, The University of Texas at Dallas, 800 West Campbell Road, Richardson, Texas 75080, USA.

*Email: cprdream@kookmin.ac.kr

# S1 Phase field microelasticity method

We describe the heterogeneous solid, i.e., polycrystalline thin film on a substrate, together with the surrounding vacuum by introducing an order parameter $\zeta$ ($\zeta = 0$ in the vacuum and $\zeta = 1$ in the thin film) following a similar approach reported in the literature[1–3]. The position dependent elastic modulus $C_{ijkl}(\mathbf{r})$ also changes the value from 0 in vacuum to $C_{ijkl}^0$ in the thin film. This sharp transition of $C_{ijkl}(\mathbf{r})$ is smoothened using a function $g(\zeta) = 0.5(1 + \frac{\tanh 2\zeta(\mathbf{r}) - 1}{2\gamma})$ to avoid mechanical instability, where $\gamma$ is the deciding parameter for the interfacial thickness between the thin film and the vacuum[4–6]. Thus, $C_{ijkl}(\mathbf{r})$ is expressed as a function of $\zeta(\mathbf{r})$ and it takes the form as $C_{ijkl}(\mathbf{r}) = C_{ijkl}^0 g(\zeta)$. The elastic energy for the inhomogeneous system can be written as

$$f_{ela} = \frac{1}{2} \int C_{ijkl}(\mathbf{r})(\epsilon_{ij} - \epsilon_{ij}^{0G})(\epsilon_{kl} - \epsilon_{kl}^{0G})dV \tag{S1}$$

We employ an iterative phase-field microelasticity solver to determine the stress field for this inhomogeneous system by considering an equivalent system with a homogeneous modulus $C_{ijkl}^0$ and an effective stress-free strain $\epsilon_{ij}^{eff}$ defined by[4,5]

$$C_{ijkl}^0\left(\epsilon_{kl}(\mathbf{r}) - \epsilon_{kl}^{eff}(\mathbf{r})\right) = C_{ijkl}(\mathbf{r})\left(\epsilon_{kl}(\mathbf{r}) - \epsilon_{kl}^{0G}(\mathbf{r})\right) \tag{S2}$$

The equivalent elastic energy considering the effective strain is written as follows:[5,7]

$$\begin{aligned}E_{el}^{eq} &= \frac{1}{2}\int C_{ijkl}^0 \epsilon_{ij}^{eff}\epsilon_{kl}^{eff}d^3r + \frac{1}{2}\int C_{ijkl}^0 E_{ij}E_{kl}d^3r \\ &- E_{ij}\frac{1}{2}\int C_{ijkl}^0\epsilon_{kl}^{eff}d^3r - \frac{1}{2}\int_{k=0}\frac{d^3r}{(2\pi)^3}\times k_i\tilde{\sigma}_{ij(\mathbf{k})}^{eff}\tilde{G}_{jk}(\mathbf{k})\tilde{\sigma}_{kl}^{eff}(\mathbf{k})k_l \\ &+ \frac{1}{2}\int_V(C_{ijmn}^0\Delta S_{mnpq}C_{pqkl}^0 - C_{ijkl}^0)\times(\epsilon_{ij}^{eff} - \epsilon_{ij}^{0G})(\epsilon_{kl}^{eff} - \epsilon_{kl}^{0G})d^3r,\end{aligned} \tag{S3}$$

where $V$ is the system volume, $G_{jk}(\mathbf{k})$ is Green's function tensor, and $k_i$ are the components of the directional vector in the Fourier space. $\Delta S_{mnpq} = [C_{mnpq}^0 - C_{mnpq}(\mathbf{r})]^{-1}$ and $\sigma_{ij}^{eff}(\mathbf{k})$ is the effective stress expressed as $\sigma_{ij}^{eff}(\mathbf{k}) = C_{ijkl}^0\int_V\epsilon_{kl}^{eff}(\mathbf{r})e^{-i\mathbf{k}\cdot\mathbf{r}}d^3r$. The effective strain can be obtained by setting the functional variation of the equivalent elastic energy with respect to the effective strain ($\frac{\delta E_{el}^{eq}}{\delta \epsilon_{ij}^{eff}} = 0$) and solving the TDGL type equation:[4–7]

$$\frac{\partial \epsilon_{ij}^{eff}}{\partial t} = -K_{ijkl}\cdot\frac{\delta E_{el}^{eq}}{\delta \epsilon_{kl}^{eff}} \tag{S4}$$

Here, the kinetic coefficient $K_{ijkl}$ is a tensor proportional to the modulus variation and assumed to be $K_{ijkl} = KS_{ijkl}^0$ for simplicity, where $S_{ijkl}^0$ is the homogeneous elastic compliance tensor and $K$ is a constant. The elastic strains and stresses are obtained from the effective strain as follows:

$$\begin{aligned}\epsilon_{ij}(\mathbf{r}) &= E_{ij} + \frac{1}{2}\int_{\mathbf{k}=0}\frac{d^3k}{(2\pi)^3}(k_i\tilde{G_{jk}} + k_j\tilde{G_{ik}})\sigma_{kl}^{\tilde{eff}}(\mathbf{k})^*k_l e^{-i\mathbf{k}\cdot\mathbf{r}} \\ \sigma_{ij}(\mathbf{r}) &= C_{ijkl}^0\left(\epsilon_{kl}(\mathbf{r}) - \epsilon_{kl}^{eff}(\mathbf{r})\right)\end{aligned} \tag{S5}$$

# S2 Derivation of stresses under plane stress state

The sum of contributions of the elastic energy and electrostrictive energy in a single grain can be expressed as

$$G = -\frac{1}{2}S_{11}\sigma_1^2 - \frac{1}{2}S_{22}\sigma_2^2 - \frac{1}{2}S_{33}\sigma_3^2 - S_{12}\sigma_1\sigma_2 - S_{23}\sigma_2\sigma_3 - S_{13}\sigma_3\sigma_1 - \frac{1}{2}S_{44}\sigma_4^2 - \frac{1}{2}S_{55}\sigma_5^2 - \frac{1}{2}S_{66}\sigma_6^2$$
$$- (Q_{13}\sigma_1 + Q_{23}\sigma_2 + Q_{33}\sigma_3)P_{Zi}^2 \quad (S6)$$

Elastic equations of state could be obtained by the minimization $\delta G/\delta\sigma_{ij} = -\mu_{jk}$. The equations of state have the form:

$$\tilde{\mu}_1 - \mu_1 = S_{11}\sigma_1 + S_{12}\sigma_2 + S_{13}\sigma_3$$
$$\tilde{\mu}_2 - \mu_2 = S_{12}\sigma_1 + S_{22}\sigma_2 + S_{23}\sigma_3$$
$$\tilde{\mu}_3 - \mu_3 = S_{13}\sigma_1 + S_{23}\sigma_2 + S_{33}\sigma_3$$
$$\tilde{\mu}_4 - \mu_4 = S_{44}\sigma_4 \quad (S7)$$
$$\tilde{\mu}_5 - \mu_5 = S_{44}\sigma_5$$
$$\tilde{\mu}_6 - \mu_6 = S_{44}\sigma_6,$$

Here, $\mu_i$ are the designations of the strains in the absence of stresses obtained from the electrostrictive contributions of the energy:

$$\mu_1 = Q_{13}P_{Zi}^2$$
$$\mu_2 = Q_{23}P_{Zi}^2$$
$$\mu_3 = Q_{33}P_{Zi}^2 \quad (S8)$$
$$\mu_4 = \mu_5 = \mu_6 = 0$$

Compatibility relation $e_{ikl}e_{jmn}(\delta^2\tilde{\mu}_{ln}/\delta x_k x_n) = 0$ leads to the conditions of constant strains $\tilde{\mu}_1$=const, $\tilde{\mu}_2$=const, and $\tilde{\mu}_6$=cont. Setting constant strains to values of spontaneous polarization in stress free, homogeneous system, we obtain

$$\tilde{\mu}_1 = \mu_1^S = Q_{13}P_0^2$$
$$\tilde{\mu}_2 = \mu_2^S = Q_{23}P_0^2 \quad (S9)$$
$$\tilde{\mu}_3 = \mu_3^S = Q_{33}P_0^2$$

Considering plane stress state $\sigma_3 = 0$, we modify Eqns. S7 to

$$S_{11}\sigma_1 + S_{12}\sigma_2 = \mu_1^S - \mu_1$$
$$S_{12}\sigma_1 + S_{22}\sigma_2 = \mu_2^S - \mu_2 \quad (S10)$$
$$S_{13}\sigma_1 + S_{23}\sigma_2 = \mu_3^S - \mu_3$$

Substituting Eqns. S8 and S9 in Eqn. S10, we obtain the unknown stress components:

$$\sigma_1 = \frac{(S_{22}Q_{13} - S_{12}Q_{23})(P_0^2 - P_{Zi}^2)}{S_{11}S_{22} - S_{12}^2}$$
$$\sigma_2 = \frac{(S_{11}Q_{23} - S_{12}Q_{13})(P_0^2 - P_{Zi}^2)}{S_{11}S_{22} - S_{12}^2} \quad (S11)$$

## S3 Calibration of Landau coefficients

The Landau coefficients for HZO are obtained by fitting polarization hysteresis function $P(E)$ with measured PE curve for 10 $nm$-thick $Hf_{0.5}Zr_{0.5}O_2$ film. $P(E)$ is the inverse of the following function:

$$E = 2a_1 P + 4a_{11} P^3 + 6a_{111} P^5 \tag{S12}$$

We obtain two sets of Landau coefficients considering first order and second order phase transitions, which are distinguished by the sign of coefficient $a_{11}$: $a_{11} < 0$ for $1^{st}$ order and $a_{11} > 0$ for $2^{nd}$ order transitions. The extracted Landau coefficients are given in **Table S1**.

| Landau coefficients | Set I ($a_{11} < 0$) | Set II ($a_{11} > 0$) |
|---|---|---|
| $a_1$ | $-2.976 \times 10^8 \; \frac{Jm}{C^2}$ | $-4.835 \times 10^8 \; \frac{Jm}{C^2}$ |
| $a_{11}$ | $-2.160 \times 10^8 \; \frac{Jm^5}{C^4}$ | $4.367 \times 10^9 \; \frac{Jm^5}{C^4}$ |
| $a_{111}$ | $1.653 \times 10^{10} \; \frac{Jm^9}{C^6}$ | $6.687 \times 10^9 \; \frac{Jm^9}{C^6}$ |

**Table S1:** Landau coefficients extracted from the measured PE curve for 10 $nm$-thick $Hf_{0.5}Zr_{0.5}O_2$ film.

The comparison of polarization hysteresis functions for two sets (Set 1: $a_{11} < 0$ and Set 2: $a_{11} > 0$) of Landau coefficients with the measured PE curve is shown in **Figure S1**. Since ferroelectric HZO is well described by first order phase transition formalism, we consider the first set ($a_{11} < 0$) of Landau coefficients for the phase field simulations.

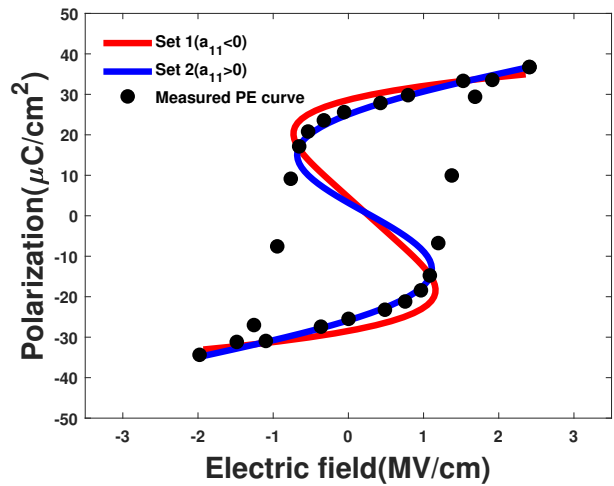

**Figure S1:** Comparison of calibrated polarization hysteresis functions for $a_{11} < 0$ (Set 1) and $a_{11} > 0$ (Set 2) with the measured PE curve. Measured PE curve reprinted (adapted) with permission from (Kim, Si Joon, et al. *ACS applied materials & interfaces* 11.5 (2019): 5208-5214)[8]. Copyright (2019) American Chemical Society.

## S4 Optimization of Landau coefficients by genetic algorithm

Two sets of optimized Landau coefficients are obtained using the genetic algorithm by considering first order and second order transitions. The optimized Landau coefficients are listed in **Table S2**.

The comparison of polarization hysteresis functions and simulated PE curves for two sets (Set 1: $a_{11} < 0$ and Set2: $a_{11} > 0$) of GA optimized Landau coefficients are presented in **Figure S2a** and **Figure S2b**, respectively. The calculated values of the

| Landau coefficients | Set I ($a_{11} < 0$) | Set II ($a_{11} > 0$) |
|---|---|---|
| $a_1$ | $-4.289 \times 10^8$ $\frac{Jm}{C^2}$ | $-4.754 \times 10^8$ $\frac{Jm}{C^2}$ |
| $a_{11}$ | $-2.242 \times 10^8$ $\frac{Jm^5}{C^4}$ | $2.848 \times 10^7$ $\frac{Jm^5}{C^4}$ |
| $a_{111}$ | $2.170 \times 10^9$ $\frac{Jm^9}{C^6}$ | $1.776 \times 10^9$ $\frac{Jm^9}{C^6}$ |

**Table S2:** Landau coefficients optimized by genetic algorithm.

objective functions are 0.00109 for Set 1 and 0.00103 for Set 2. Here, we consider the first set ($a_{11} < 0$) of Landau coefficients describing first order phase transition for the phase field simulations.

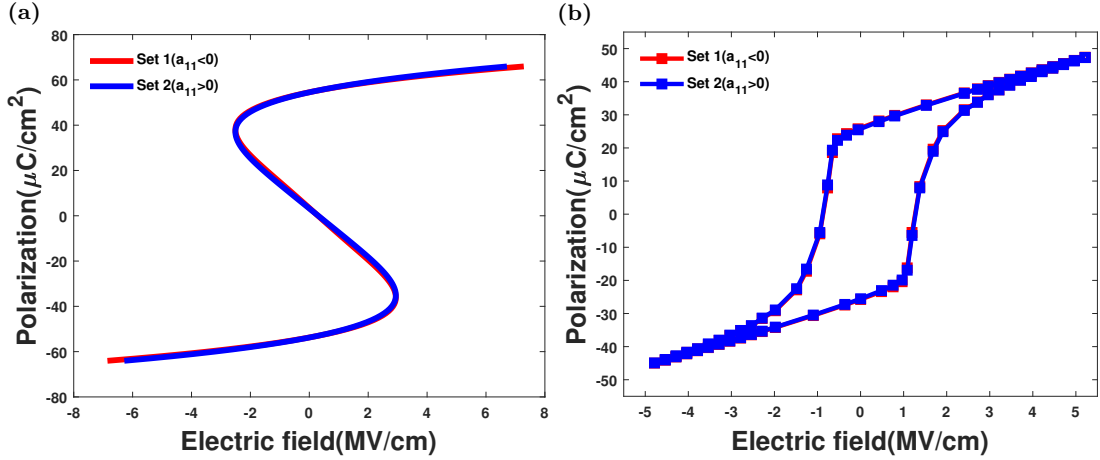

**Figure S2:** (a) Comparison of GA optimized polarization hysteresis functions for $a_{11} < 0$ (Set 1) and $a_{11} > 0$ (Set 2). (b) Comparison of simulated PE curves using the same two sets of GA optimized Landau coefficients.

## S5   Estimation of spontaneous polarization and dielectric permittivity of HZO from GA optimized effective Landau coefficients

The single-domain ground state property of the HZO including the spontaneous polarization and dielectric permittivity are calculated from the GA optimized Landau coefficients. The spontaneous polarization $P_0$ is obtained from Landau coefficients using the following expression:[9]

$$P_0 = \pm \sqrt{-\frac{a_{11}\left(1 + \sqrt{1 - \frac{3a_1 a_{111}}{a_{11}^2}}\right)}{3a_{111}}} \quad (S13)$$

The dielectric permittivity $\epsilon_r$ is calculated from the expression[10]

$$\epsilon_r = \frac{1}{3\sqrt{3}} \frac{P_0}{\epsilon_0 E_c}, \quad (S14)$$

where $E_c$ is the coercive field estimated from the given expression:[9]

$$E_c = \frac{8}{25} \sqrt{\frac{|a_{11}|^5}{5a_{111}^3}} \sqrt{1 + \sqrt{1 - \frac{5a_1 a_{111}}{a_{11}^2}}} \left(1 - \frac{5a_1 a_{111}}{a_{11}^2} + \sqrt{1 - \frac{5a_1 a_{111}}{a_{11}^2}}\right) \quad (S15)$$

## S6 Normalization of applied electric fields

For better comparison of the domain dynamics during switching in phase field simulations and experimental observations, the applied electric fields are normalized with respect to corresponding switching voltages. The switching regions are estimated from the fraction of up and down polar domains in the domain structures at different applied fields. The reversal of polarization from up to down domains occurs from $-1.2 \frac{MV}{cm}$ to $-1.8 \frac{MV}{cm}$ in experiments and from $-0.75 \frac{MV}{cm}$ to $-2.25 \frac{MV}{cm}$ in simulations. Similarly, polarization reversal from down to up domains occurs from $1.2 \frac{MV}{cm}$ to $1.8 \frac{MV}{cm}$ in experiments and from $0.75 \frac{MV}{cm}$ to $2.25 \frac{MV}{cm}$ in simulations. The normalized applied field $E^*$ is calculated using the following expressions:

$$E^* = \frac{5}{3}\left(1 - \frac{3}{5|E_{exp}|}\right)E_{exp}, \tag{S16}$$

$$E^* = \frac{4}{7}\left(1 + \frac{1}{|E_{pfm}|}\right)E_{pfm}, \tag{S17}$$

where, $E_{exp}$ and $E_{pfm}$ are the applied electric fields in experiments and simulations, respectively. The electric fields applied on the thin film in simulations and experiments corresponding to the values of normalized applied fields for domain structures illustrated in **Figure 4** are listed in **Table S3**.

| $E^*$ | $E_{exp}(MV/cm)$ | $E_{pfm}(MV/cm)$ |
|---|---|---|
| $-1.2$ | $-1.3$ | $-1.0$ |
| $-1.3$ | $-1.4$ | $-1.3$ |
| $-1.5$ | $-1.5$ | $-1.6$ |
| $-1.8$ | $-1.7$ | $-2.2$ |
| $1.2$ | $1.3$ | $1.0$ |
| $1.3$ | $1.4$ | $1.3$ |
| $1.5$ | $1.5$ | $1.6$ |
| $1.8$ | $1.7$ | $2.2$ |

**Table S3:** Applied electric fields in simulations and experiments corresponding to values of normalized applied fields for domain structures in Figure 4

## S7 Simulation of polarization profiles across a 180° domain wall and switching curves in a bulk single crystal using calibrated effective Landau coefficients and corresponding Landau coefficients

The calibrated effective Landau coefficients ($a_1 = -2.976 \times 10^8 \frac{Jm}{C^2}$, $a_{11} = -2.160 \times 10^8 \frac{Jm^5}{C^4}$, and $a_{111} = 1.653 \times 10^{10} \frac{Jm^9}{C^6}$) are used to simulate polarization profiles across a 180° domain wall and polarization switching in a bulk single crystal (two-dimensional). Further, we perform the simulations using Landau coefficients considering elastic energy and compare both results. The Landau coefficients estimated from the calibrated effective coefficients are: $\alpha_1 = -2.749 \times 10^8 \frac{Jm}{C^2}$, $\alpha_{11} = -4.933 \times 10^8 \frac{Jm^5}{C^4}$, and $\alpha_{111} = 1.653 \times 10^{10} \frac{Jm^9}{C^6}$. The polarization profiles for the two domain system for both sets of coefficients are compared in **Figure S3**. The values of steady-state polarizations in 180° domains are matching for the effective Landau coefficients and corresponding Landau coefficients. The comparison of switching curves in bulk single crystal for both sets of coefficients are given in **Figure S4**. The PE curves are also similar for the effective Landau coefficients and corresponding Landau coefficients except for some discrepancies in the coercive field. This shows good agreement of the strain-polarization coupling with the free energy coefficients.

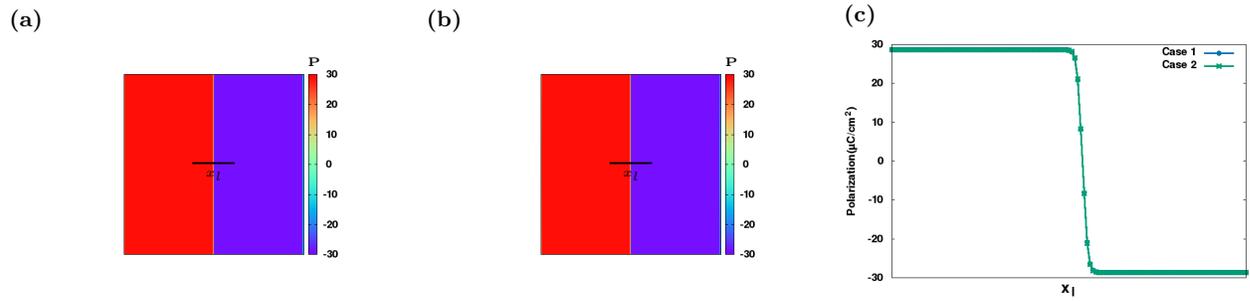

**Figure S3:** Polarization profiles across a 180° domain wall simulated using (a) calibrated effective Landau coefficients (Case 1) and (b) corresponding Landau coefficients (Case 2). (c) Comparison of the polarization profiles (1D) for Case 1 and Case 2 along a line $x_l$ drawn across the domain walls.

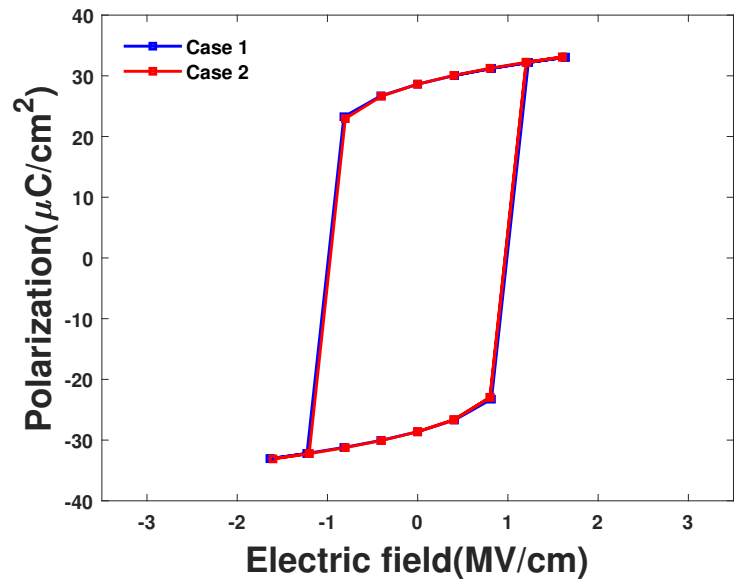

**Figure S4:** Comparison of PE curves for single crystal simulated using calibrated effective Landau coefficients in the absence of elastic interactions (Case 1) and corresponding Landau coefficients considering elastic interactions (Case 2).

## S8  Estimation of susceptibility of the non-ferroelectric phase ($\chi_d$)

To study the impact of the ferroelectric phase fraction of HZO thin films on switching characteristics, the susceptibility of the non-ferroelectric phase is obtained from the measured PE curve of 10 $nm$-thick $HfO_2$ film. PE curves are generated using a phase field model for non-ferroelectric columnar thin films with bulk free energy expressed as follows:

$$f_{bulk} = \sum_{i=1}^{N} h(\phi_i) f_{Di}, \tag{S18}$$

where $N$ is the total number of grains. We estimate the value of dielectric susceptibility at which the simulated PE curve matches the experimentally calculated PE curve for non-ferroelectric hafnia thin film. The fit between simulated and measured PE curves are demonstrated in **Figure S5** and the estimated value of $\chi_d$ is $2.108 \times 10^{-10}\ \frac{C^2}{Jm}$.

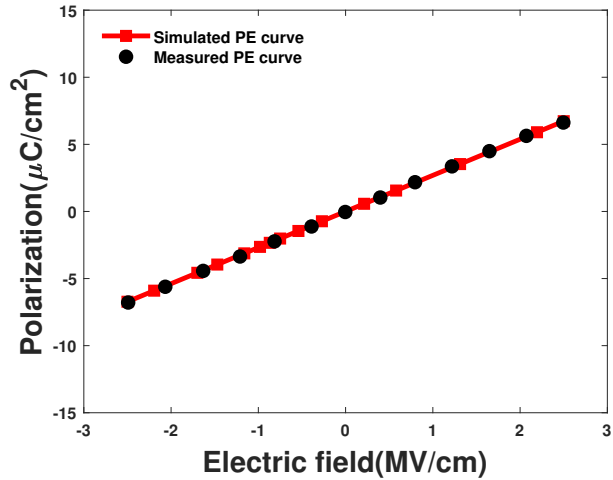

**Figure S5:** Comparison of simulated PE curve for non-ferroelectric columnar thin film and experimentally measured curve for 10 $nm$-thick $HfO_2$ film. Measured PE curve reprinted (adapted) with permission from (Kim, Si Joon, et al. *ACS applied materials & interfaces* 11.5 (2019): 5208-5214)[8]. Copyright (2019) American Chemical Society.

## S9  Comparison of PE curves of [001] textured film and single crystalline HZO.

We compare the PE curve for [001] textured film with the PE curve for single crystalline HZO. From **Figure S6**, it is evident that there are discrepancies between the switching characteristics of [001] textured thin film and single crystalline HZO. Even though the remnant polarizations for both curves are similar, the coercive field and polarization switching slopes are different. This mismatch in coercive field and polarization switching slope can be due to the reduced degree of preferred orientation in the textured thin film. When the orientations of grains in the textured film deviates from the preferred orientation, the non-polar contributions of the free energy $\left(1/2\chi_f(P_{Xi}^2 + P_{Yi}^2)\right)$ affects the polarization switching modifying coercive field and switching slope. Because of its quadratic form, the non-ferroelectric contribution cannot significantly influence the switching curve in the absence of any external field. Therefore, the remnant polarizations for both [001] textured thin film and single crystalline HZO remain similar.

## S10  Effect of grain size on switching characteristics

Simulations are performed on 10 $nm$-thick columnar films of different grain sizes ($D = 60, 90$ and $120\ nm$) to investigate the impact of grain size on switching dynamics. **Table S4** provides the simulation parameters used for the study. The grain size dependent

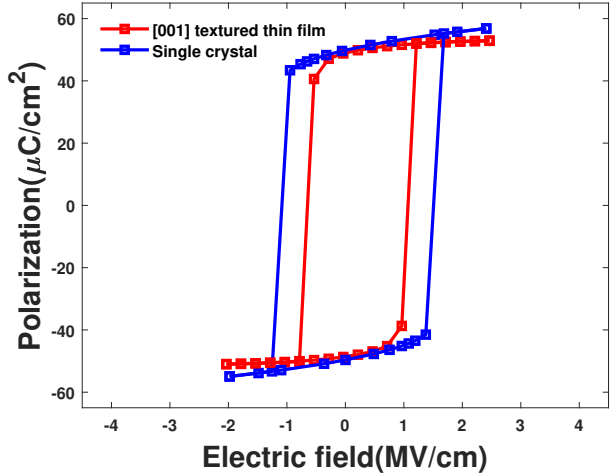

**Figure S6:** Comparison of simulated PE curve for [001] textured thin films with the PE curve of single crystalline HZO.

behaviour of PE curves are illustrated in **Figure S7**. It shows that the switching characteristics do not significantly change with the variation in grain sizes. With an increase in grain size, remnant polarization increases, but the coercive field remains almost the same ($\sim 1.1\ \frac{MV}{cm}$). Thin films with grain sizes of $60, 90$ and $120\ nm$ exhibit $P_r$ values of $\sim 25.86\ \frac{\mu C}{cm^2}$, $\sim 28.21\ \frac{\mu C}{cm^2}$, and $\sim 29.66\ \frac{\mu C}{cm^2}$, respectively.

The reduced remnant polarization in thin films with finer grains can be attributed to the increase in the proportion of grain boundaries with the decrease in grain size. The grain boundaries may distort ferroelectric switching because of their paraelectric nature. Our simulations certainly indicate an enhancement in ferroelectricity in thin films with coarser grains. But, we have not considered the change of ferroelectric orthorhombic phase to non-ferroelectric monoclinic phase with an increase in grain size as reported in experiments [11,12]. Therefore, further systematic studies are required to understand the influence of grain size on ferroelectric properties of HZO thin films.

| Parameters | |
| --- | --- |
| Thin film surface area ($\Delta X \times \Delta Y$) | $1280 \times 1280, 2240 \times 2240, 2688 \times 2688\ nm^2$ |
| Thin film thickness ($t_Z$) | $10\ nm$ |
| Effective Landau coefficient ($a_1$) | $-4.289 \times 10^8\ \frac{Jm}{C^2}$ |
| Effective Landau coefficient ($a_{11}$) | $-2.242 \times 10^8\ \frac{Jm^5}{C^4}$ |
| Effective Landau coefficient ($a_{111}$) | $2.170 \times 10^9\ \frac{Jm^9}{C^6}$ |
| Background dielectric susceptibility ($\chi_f$) | $4.019 \times 10^{-10}\ \frac{C^2}{Jm}$ |
| Gradient energy coefficient ($G_{11}$) | $5.066 \times 10^{-10}\ \frac{Jm^3}{C^2}$ |
| Average grain size ($D$) | $60, 90, 120\ nm$ |
| Number of grains ($N$) | $480, 492, 496$ |

**Table S4:** Simulation parameters used to investigate the effect of grain size in thin films with columnar grain morphology.

To investigate the effect of grain size on switching dynamics, we use $20\ nm$-thick films with an equiaxed grain morphology and different grain sizes ($D = 20, 30$ and $40\ nm$). GA optimized Landau coefficients are used in the simulations and the parameters are listed in **Table S5**. The comparison of simulated PE curves for thin films with different grain sizes is illustrated in **Figure S8**.

Following the trend seen in columnar thin films, remnant polarization increases from $\sim 22.80\ \frac{\mu C}{cm^2}$ to $\sim 25.65\ \frac{\mu C}{cm^2}$ and the coercive field remains essentially the same at $\sim 1.3\ \frac{MV}{cm}$, when the grain size is varied from 20 to 30 $nm$. A further increase in grain size to 40 $nm$ also increases the remnant polarization to $\sim 28.16\ \frac{\mu C}{cm^2}$, but the coercive field decreases to $\sim 1.2\ \frac{MV}{cm}$. This variation in switching characteristics can be explained by the change in grain morphology from an equiaxed to a columnar structure with an increase in grain size. When the average grain size increases beyond thin film thickness, the equiaxed grain structure tends

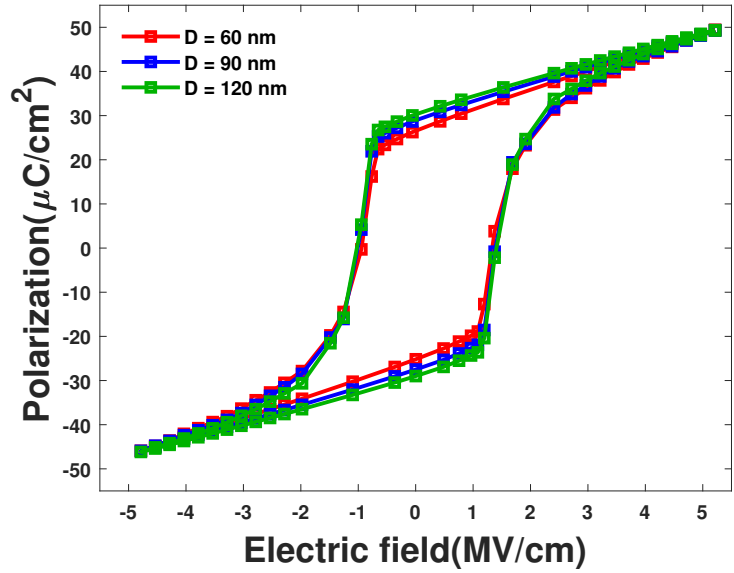

**Figure S7:** Effect of grain size on switching characteristics. Comparison of simulated PE curves for columnar thin films ($t_F = 10\ nm$) with different grain sizes ($D = 60, 90, 120\ nm$).

| Parameters | |
|---|---|
| Thin film surface area ($\Delta X \times \Delta Y$) | $480 \times 480, 800 \times 800, 1200 \times 1200\ nm^2$ |
| Thin film thickness ($t_Z$) | $20\ nm$ |
| Effective Landau coefficient ($a_1$) | $-4.289 \times 10^8\ \frac{Jm}{C^2}$ |
| Effective Landau coefficient ($a_{11}$) | $-2.242 \times 10^8\ \frac{Jm^5}{C^4}$ |
| Effective Landau coefficient ($a_{111}$) | $2.170 \times 10^9\ \frac{Jm^9}{C^6}$ |
| Background dielectric susceptibility ($\chi_f$) | $4.019 \times 10^{-10}\ \frac{C^2}{Jm}$ |
| Gradient energy coefficient ($G_{11}$) | $5.066 \times 10^{-10}\ \frac{Jm^3}{C^2}$ |
| Average grain size ($D$) | $20, 30, 40\ nm$ |
| Number of grains ($N$) | $427, 620, 756$ |

**Table S5:** Simulation parameters to study the effect of grain size in thin films with equiaxed grain morphology.

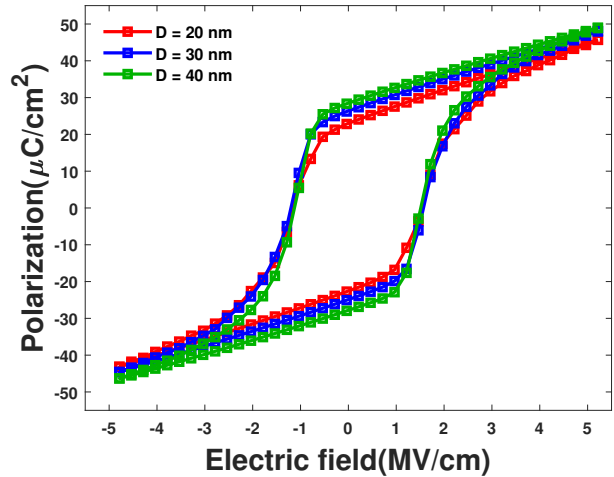

**Figure S8:** Effect of grain size on switching characteristics. Comparison of simulated PE curves for 20 $nm$-thick equiaxed films with different grain sizes ($D = 20, 30, 40\ nm$).

to become columnar. This is evident from the side views ($ZX$ plane) of the simulated microstructure of the thin films shown in **Figure S9**.

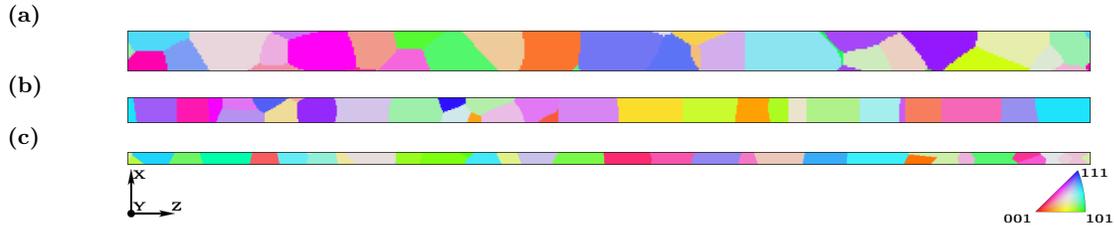

**Figure S9:** Side views ($ZX$ plane) of the simulated microstructure of 20 $nm$-thick equiaxed films with different grain sizes: (a) 20, (b) 30, and (c) 40 $nm$.

The increase in remnant polarization with the increase in grain size from 20 to 30 $nm$ is due to the decreasing proportion of grain boundaries when the grain size increases. Whereas, the increase in remnant polarization and decrease in the coercive field with a further increase in grain size to 40 $nm$ can be attributed to the change in grain morphology of the thin film.


\clearpage
\end{document}